\documentclass[aps,10pt,tightenlines,superscriptaddress,nofootinbib]{revtex4}
\usepackage{graphicx}
\usepackage{xcolor}
\usepackage{caption}
\usepackage{subcaption}
\newcommand{\bea}{\begin{eqnarray}}
\newcommand{\eea}{\end{eqnarray}}
\usepackage{multirow}
\usepackage{amsmath,amssymb}  
\usepackage{bm}
\usepackage{float}
\usepackage{graphics}
\usepackage{slashed}
\usepackage{comment}
\usepackage[normalem]{ulem}
\usepackage{verbatim}

\usepackage[unicode=true,hidelinks,colorlinks,linkcolor=black,citecolor={green!70!black},urlcolor={blue!70!black}]{hyperref}

\begin{document}

\title{Fully charmed tetraquark production at the LHC experiments}

\author{Ilia Belov}
\email[]{ilia.belov@ge.infn.it}
\affiliation{INFN, Sezione di Genova, Via Dodecaneso 33, I-16146 Genova, Italy}

\author{Alessandro Giachino}
\affiliation{Departamento de F\'{\i}sica Te\'orica and IFIC,
Centro Mixto Universidad de Valencia-CSIC, Parc Científic UV, C/ Catedrático José Beltrán, 2, 46980 Paterna, Spain}
 
\author{Elena Santopinto}
\affiliation{INFN, Sezione di Genova, Via Dodecaneso 33, I-16146 Genova, Italy}

\begin{abstract}
We develop the formalism for production of a fully heavy tetraquark and apply it to the calculation of $pp\to T_{4c}+X$ cross-sections. We demonstrate that the production cross-section of a fully heavy tetraquark, even if it is a diquark-antidiquark cluster, can be obtained in the meson-like basis, for which the spin-color projection technique is well established. Prompted by the recent LHCb, ATLAS and CMS data, we perform a pQCD calculation of ${\cal O}(\alpha_s^5)$ short-distance factors in the dominant channel of gluon fusion, and match these to the four-body $T_{4c}$ wave functions in order to obtain the unpolarized $T_{4c}(0^{++},1^{+-},2^{++})$ cross-sections. The novelty in comparison with the recently published article~\cite{Feng:2023agq} lies in the fact that we predict the absolute values as well as the $d\sigma/dp_T$ spectra in the kinematic ranges accessible at the ongoing LHC experiments. From the comparison with the signal yield at LHCb we derive the constraints on the $\Phi\cdot\text{Br}(J/\psi\,J/\psi)$~(reduced wave function times branching) product for the $T_{4c}$ candidates for $X(6900)$ and observe that $X(6900)$ is compatible with a $2^{++}(2S)$ state.
\end{abstract}

\maketitle

\section{Introduction}
The 2020 LHCb discovery of a narrow structure near $6900\,\rm MeV$ in the di-$J/\psi$ invariant mass spectrum~\cite{LHCb:2020bwg} inaugurated the experimental era of the fully heavy tetraquark states. The mass and width of this state, named $X(6900)$, have been measured as $M=(6886 \pm 11 \pm 11 )\; \text{MeV}$ and $\Gamma=(168 \pm 33 \pm 69) \;\text{MeV}$~\cite{LHCb:2020bwg}. In addition to the $X(6900)$ state, the LHCb collaboration also reported evidence of two additional structures in the same decay channel, peaking at approximately 6400~MeV and 7200~MeV. LHCb also measured the production ratio $\mathcal{R}$, defined as the number of $X(6900)$ signal events over the total number of di-$J/\psi$ events, $\mathcal{R}=N(X(6900)\to J/\psi\,J/\psi)):N_\textrm{tot}(J/\psi\,J/\psi)$, where $N_\textrm{tot}(J/\psi\,J/\psi)$ is  the sum of the background and signal \cite{LHCb:2020bwg}. The background represents  $J/\psi$-pairs, either promptly produced or originating as feed-down components from higher charmonium states.

Later,  ATLAS also observed a significant excess of events over the total background in the di-$J/\psi$ mass spectrum. As well as confirming the $X(6900)$, ATLAS  reported a broad structure starting from the di-$J/\psi$ threshold with a global statistical significance far exceeding $5\,\sigma$. ATLAS also explored the $J/\psi\, \psi(2S)$ decay channel and reported an additional peak not seen in the di-$J/\psi$ mode, with statistical significance larger than  $3\,\sigma$ \cite{ATLAS:2023bft}.

Recently, CMS investigated the di-$J/\psi$ mass spectrum and confirmed the previously discovered $X(6900)$ state~\cite{CMS:2023owd}. In that work~\cite{CMS:2023owd}, the CMS collaboration found two new states,  namely the $X(6600)$ and $X(7300)$ states, with $6.9$ and  $4.7$ standard deviations, respectively.

Up to 2023, investigation of the fully heavy tetraquark production was guided by implementation of the ${\cal O}(\alpha_s^4)$ matrix elements squared for the hard scattering reaction~\cite{Berezhnoy:2011xy,Berezhnoy:2011xn,Maciula:2020wri,Zhu:2020xni,Becchi:2022dps} and, moreover,  normalization of the model-dependent parameters that describe the four-quark hadronization relied on qualitative considerations.

In fact, the first courageous attempts to estimate the fully heavy tetraquark production cross-sections were made in 2011~\cite{Berezhnoy:2011xy,Berezhnoy:2011xn}. In those studies the authors calculated the $gg\to |cc\rangle_{\bar 3}\otimes |\bar{c}\bar{c}\rangle_{3}$  matrix elements and used Pythia  to generate a $p_T$ smearing induced by the Initial State Radiation (ISR). Using duality relations, they predicted a qualitative shape of $\sigma(pp\to T_{4c}\to J/\psi\,J/\psi)$ distributions in the approximation of a point-like diquark-antidiquark coupled to a $T_{4c}$.

In 2021, the  $T_{4c}$ production mechanism was studied at $\mathcal O (\alpha_s^4)$ order by means of the $k_T$-factorization approach describing hadronization in the Color Evaporation Model (CEM)~\cite{Maciula:2020wri}. Nevertheless, in CEM the probability of $c\bar c c\bar c \to T_{4c}$ transition, which is crucial to obtaining some quantitative predictions, does not come from first principles rather it was estimated by trying to reproduce the LHCb signal-to-background ratio measured in~\cite{LHCb:2020bwg}. Both Single Parton Scattering (SPS) and Double Parton Scattering (DPS) mechanisms were taken into account in estimating the di-$J/\psi$ background and $T_{4c}$ signal. One of the main observations made in~\cite{Maciula:2020wri} is that in CEM the DPS contribution exhibits two order in magnitude enhancement in respect with the SPS contribution even for $c\bar{c}c\bar{c}$ invariant masses in the vicinity of the $T_{4c}$ position. The authors of~\cite{Maciula:2020wri}, however, do not make a conclusion on the DPS dominance, arguing that there is no reliable approach for the DPS formation mechanism of tetraquarks at present. In this matter, we share the same point of view, and therefore throughout the article, we consider $T_{4c}$ production in the SPS interaction, in which the produced quarks and antiquarks may be trusted to be enough space-time correlated.

In~\cite{Zhu:2020xni} the low $p_T$ limit, $p_T \ll m_{4c}$, of the inclusive $\sigma(pp\to T_{4c}+X)$ cross-section was estimated from the $\sigma(pp\to T_{4c})$ cross-section with the help of Transverse-Momentum-Dependent (TMD) factorization. Unfortunately, this  $p_T$ limit is out of range for ATLAS and CMS, while for LHCb it is still too low to collect enough statistics.

A few estimations of the production cross-section at the threshold, $\sigma(pp\to T_{4c})$, were made in~\cite{Carvalho:2015nqf,Abreu:2023wwg,Becchi:2022dps}. The problem with the production calculations at threshold is that the tetraquark is produced without a transverse momentum, making it difficult to observe at the LHC. Indeed, the ${\cal O}(\alpha_s^4)$ scattering process produces a tetraquark with zero transverse momentum  ($\sigma_{2\to 1}\sim \delta(p_T)$), as required by momentum conservation. This means that, unless one introduces the nonzero transverse momentum component in the initial state partons in an effective way (consider, for example, the ISR simulation or the $k_T$-factorization scheme), one has to study the inclusive production process.

To date, only one theoretical calculation of $T_{4c}$ inclusive hadronic cross-section has been made~\cite{Feng:2023agq}; this calculation is based on NRQCD at the lowest order in $v$~(velocity) expansion. Unfortunately, it will be challenging to compare their results with the future experimental data, as they did not account for the experimental cuts in the calculation of the hadronic cross-section. For this reason, the goal of the present article is to obtain the  $T_{4c}$ cross-sections that can be compared in the future with the cross-sections at the LHC experiments.

Now, we develop the production formalism for an $S$-wave fully heavy tetraquark based on a factorization scheme which employes the tetraquark four-body wave function. We present the calculation of the $\sigma(pp\to T_{4c}+X)$ cross-section at the  ${\cal O}(\alpha_s^5)$ order within the Color Singlet Model (CSM) by means of a projection technique different from the one employed in~\cite{Feng:2023agq}. The authors of~\cite{Feng:2023agq} apply the amplitude projection onto the diquark-antidiquark cluster, $|c c\rangle \otimes |\bar c\bar c\rangle$, with the help of covariant projectors derived previously in~\cite{Feng:2020riv}.\,\footnote{The authors of Refs.~\cite{Feng:2023agq,Feng:2020riv} apply in~\cite{Feng:2023agq} the projectors in the diquark-antidiquark basis proposed earlier in~\cite{Feng:2020riv}. In our opinion, this projection technique  was only partially explained  in~\cite{Feng:2020riv}, and no further details were provided in~\cite{Feng:2023agq}. Indeed, we tried to reproduce the results of~\cite{Feng:2023agq}, but failed because we could not find the recipe for the implementation of the projectors of Eq.~(10) of Ref.~\cite{Feng:2020riv}.  Therefore, it might be helpful if the authors of~\cite{Feng:2020riv,Feng:2023agq} could provide more details on the derivation and implementation of their projection technique in the diquark-antidiquark basis.} By contrast, we perform the amplitude projection onto the  $T_{4c}$ in the $|c\bar c\rangle\otimes|c\bar c\rangle$ basis, for which the spin-color projection techniques are well established~\cite{Guberina:1980dc,Kuhn:1979bb}. With this approach, a production amplitude of the diquark-antidiquark cluster forming $T_{4c}$ can be decomposed into the sum of a few amplitudes in the meson-like basis. This technique of basis transformation was proposed in Refs.~\cite{Becchi:2020uvq,Becchi:2020mjz}, in which the $T_{4c}$ and $T_{4b}$ partial decay widths were estimated.  
Concerning the production formalism, the use of the covariant projectors onto the $|c\bar c\rangle$-states, which are widely applied to the study of quarkonium production, validates the reliability of our results and greatly reduces the computational cost. In our calculation, the IR divergencies are regularized with a $p_T$ cut-off, analogous to the IR regularization of the SPS NLO$^*$ contribution to $\sigma(pp\to J/\psi\,J/\psi+X)$ performed in~\cite{Lansberg:2014swa,Likhoded:2016zmk}.

We emphasize that we carry out an accurate pQCD calculation of the production matrix elements without utilizing any of the multipurpose Monte Carlo event generators developed in the high-energy physics community. It is worth noting that our formalism can be applied in the future also to the $bb\bar{b}\bar{b}$  and $bc\bar{b}\bar{c}$ tetraquarks.

The article is organized as follows. In Section~\ref{sec:phenomen}, we present the formalism of $T_{4c}\left(0^{++},1^{+-},2^{++}\right)$ production: we construct the production amplitudes in the meson-like basis with the relativistic degrees of freedom totally integrated out. In Section~\ref{sec:calc}, we apply this formalism to the  $gg\to c\bar c c \bar c g$ scattering process,  calculate the gauge invariant matrix elements for the $T_{4c}+g$ final state and discuss the impact of the ISR contribution. In Section~\ref{sec:res}, we present the hadronic cross-sections in the kinematic ranges relevant to the ongoing experiments at LHC. In Section~\ref{sec:yields}, we estimate the $T_{4c}$ yields in the di-$J/\psi$  mass spectrum at LHCb and we derive constraints to discriminate among the possible $T_{4c}$ candidates for $X(6900)$. In Section~\ref{sec:disc}, we figure out which di-$J/\psi$ structures can be interpreted as genuine resonances of the $T_{4c}$ family. In Section~\ref{sec:concl}, we report our results.

\section{Phenomenology of $T_{4c}$ production}
\label{sec:phenomen}
In the collinear approach, the leading-order process of the $T_{4c}$ production with experimentally observable transverse momentum necessarily implies the $2\to 2$ scattering process, which occurs at the ${\cal O}(\alpha_s^5)$ order.
Moreover,  in the LHC energy range, the gluon-gluon luminosity density dominates over the  quark-gluon and quark-antiquark one. 
In the dominant gluon-gluon channel one needs to consider a process with at least one real gluon in the final state: 
\begin{equation}
g (k_1)\, g (k_2) \to T_{4c}(P) \,+\,g (k_3)\,,
\label{proc}
\end{equation}
where a final state hard gluon recoils against the heavy resonance $T_{4c}$, thus leading to a measurable transverse momentum.

For tetraquark production, we adopt the CSM, which was originally employed for charmonium production. In this model the $c \bar{c}$-pair is projected onto the color-singlet Fock state, and the quantum numbers of the $c \bar{c}$-pair do not evolve between the production of the heavy-quark pair and its formation into the bound state. The only difference in the case of a fully charmed $c \bar{c} c \bar{c}$ tetraquark is that now each $c \bar{c}$-pair is projected onto $|c \bar{c} \rangle$-configuration with a particular spin and color. As we will show later, the two projected $|c \bar{c} \rangle$-configurations are coupled to a color-singlet $|cc\rangle\otimes |\bar{c}\bar{c}\rangle$ configuration with the spin of $T_{4c}$.

Another possibility is to study the color-octet production mechanism. In this approach the $|cc\rangle\otimes |\bar{c}\bar{c}\rangle$ Fock states are produced in color-octet, and their hadronization to $T_{4c}$ is associated with a soft gluon(s) emission necessary to bring the four-quark system to the colorless state. Of course, before definite predictions can be made, numerical values for the hadronization probabilities must be known. Given the absence of those, we do not construct the relevant color-octet Fock states in this article, although we acknowledge that this work may be feasible in the future. Thus, in the rest of the text, we mean CSM by default.

Let us recall that the $0^{++}$ tetraquark is a superposition of two diquark-antidiquark configurations, namely the color $[\bar 3\otimes 3]$ with spin $[1\otimes 1]$, and the color $[6\otimes \bar 6]$ with spin $[0\otimes 0]$, whose weights, $\alpha$ and $\beta$, are normalized according to $\alpha^2+\beta^2=1$, and since the flavor wave function of two identical quarks is symmetric, the  Pauli principle requires that a color-triplet diquark has spin 1, while a color-sextet diquark has spin 0. As a result,  the $T_{4c}\left(0^{++}\right)$ is a superposition of   $\alpha\,\big||cc\rangle_{\bar{3}}^{\;1} \otimes |\bar{c}\bar{c}\rangle_{3}^{\;1} \big\rangle_{1}^{\;0} + \beta\,\big||cc\rangle_{\bar{6}}^{\;0} \otimes |\bar{c}\bar{c}\rangle_{6}^{\;0} \big\rangle_{1}^{\;0}$, where the subscripts denote the dimension of the color representations, and the superscripts denote the spin, while the $T_{4c}\left(1^{+-}\right)$ is $ \left| |cc\rangle_{\bar{3}}^{\;1} \otimes |\bar{c}\bar{c}\rangle_{3}^{\;1} \right\rangle_{1}^{\;1}$, and the $T_{4c}\left(2^{++}\right)$ is $\left| |cc\rangle_{\bar{3}}^{\;1} \otimes |\bar{c}\bar{c}\rangle_{3}^{\;1} \right\rangle_{1}^{\;2}$.

In the framework of the four-body approach, the tetraquark with a given spin $J$ is a four-quark nonrelativistic object coupled through the wave function $\Psi^{(J)}({\bf q_1},{\bf q_2},{\bf q_{12}})$ with dimension of mass to the power of $9/2$ and normalized to unity
\begin{align}\nonumber
& \Psi^{(0)}({\bf q_1},{\bf q_2},{\bf q_{12}}) \equiv \Psi^{(0)}_{[\bar 3\otimes 3]+[6\otimes\bar 6]}({\bf q_1},{\bf q_2},{\bf q_{12}}),\\ 
& \Psi^{(1)}({\bf q_1},{\bf q_2},{\bf q_{12}}) \equiv \Psi^{(1)}_{[\bar 3\otimes 3]}({\bf q_1},{\bf q_2},{\bf q_{12}}),\\ \nonumber
& \Psi^{(2)}({\bf q_1},{\bf q_2},{\bf q_{12}}) \equiv \Psi^{(2)}_{[\bar 3\otimes 3]}({\bf q_1},{\bf q_2},{\bf q_{12}}).
\end{align}

The $S$-wave production amplitude can be obtained by convoluting the wave function with the hard process matrix element, $\:\widetilde{\cal M}\left({\bf q_1},{\bf q_2},{\bf q_{12}}\,\big|\, gg\to \left[c\bar cc\bar c\right](J)+g\right)$ in the colliding partons center-of-mass frame. As the wave function is assumed to be rapidly damped in the relative momenta, the scattering matrix element dependence on $({\bf q_1},{\bf q_2},{\bf q_{12}})$ can be dropped~\cite{Guberina:1980dc,Kuhn:1979bb}: 
\begin{multline}\label{main}
{\cal A}(gg \to T_{4c}(J)+g) = I_\textrm{ph.sp.}\int \frac{d {\bf q_1}}{(2\pi)^3}\frac{d {\bf q_2}}{(2\pi)^3}\frac{d {\bf q_{12}}}{(2\pi)^3}\Psi^{(J)}({\bf q_1},{\bf q_2},{\bf q_{12}})\:\widetilde{\cal M}\left({\bf q_1},{\bf q_2},{\bf q_{12}}\,\big|\, gg\to \left[c\bar cc\bar c\right](J)+g\right)\approx \\ 
\approx I_\textrm{ph.sp.}\Psi^{(J)}(0,0,0)\:{\cal M}\left(gg\to \left[c\bar cc\bar c\right](J)+g\right).
\end{multline}  
On the right-hand side of Eq.~\eqref{main} we used the approximation of neglecting the relative motion. In this approximation, the production amplitude reduces to the product of the wave function at the origin and the scattering matrix element projected onto the $|c\bar c c\bar c\rangle$ Fock state with a given spin $J$.
The factor $I_\textrm{ph.sp.}$ is a phase space factor consistent with the wave function; this will be explained later in~Section~\ref{sec:calc}.

The approximation of neglecting the relative motion used in Eq.~\eqref{main} also allows one to move safely from the diquark-antidiquark basis to the meson-like basis in the calculation of ${\cal M}\left(gg\to \left[c\bar cc\bar c\right](J)+g\right)$. As long as the relative motion is neglected,  the total amplitude is insensitive to the dynamics of heavy quarks in the $T_{4c}$ rest frame. This means that the production amplitudes calculated with the factorization given by the right-hand side of Eq.~\eqref{main}  are independent of the choice of the basis for building the $|c\bar cc\bar c\rangle$ Fock states.

The change of basis in ${\cal M}\left(gg\to \left[c\bar cc\bar c\right](J)+g\right)$ can be performed by using the  Fierz transformation in spin and color space. The relations between the diquark-antidiquark basis and the meson-like basis have been discussed in previous studies~\cite{Becchi:2020uvq,Becchi:2020mjz}. Here, we report them for completeness (the subscripts denote the dimension of the color representations, and the superscripts denote the spin): 
\begin{align}\notag 
&\big||cc\rangle_{\bar{3}}^{\;1} \otimes |\bar{c}\bar{c}\rangle_{3}^{\;1} \big\rangle_{1}^{\;0} = -\frac{1}{2}\left(\sqrt{\frac{1}{3}} \big| |c \bar{c} \rangle^{\;1}_{1}\otimes |c\bar{c}\rangle^{\;1}_{1} \big\rangle^{\;0}_{1}   -\sqrt{\frac{2}{3}} \big| |c \bar{c} \rangle^{\;1}_{8}\otimes |c\bar{c}\rangle^{\;1}_{8} \big\rangle^{\;0}_{1}  \right) + 
\\
\label{fierz_spin0}
&\hspace{55ex} +\frac{\sqrt{3}}{2} \left(\sqrt{\frac{1}{3}} \big| |c \bar{c} \rangle^{\;0}_{1}\otimes |c\bar{c}\rangle^{\;0}_{1} \big\rangle^{\;0}_{1}   -\sqrt{\frac{2}{3}} \big| |c \bar{c} \rangle^{\;0}_{8}\otimes |c\bar{c}\rangle^{\;0}_{8} \big\rangle^{\;0}_{1} \right),
\\ 
\notag
&\big||cc\rangle_{\bar{6}}^{\;0} \otimes |\bar{c}\bar{c}\rangle_{6}^{\;0} \big\rangle_{1}^{\;0} = \frac{\sqrt{3}}{2}\left(\sqrt{\frac{2}{3}} \big| |c \bar{c} \rangle^{\;1}_{1}\otimes |c\bar{c}\rangle^{\;1}_{1} \big\rangle^{\;0}_{1}   + \sqrt{\frac{1}{3}} \big| |c \bar{c} \rangle^{\;1}_{8}\otimes |c\bar{c}\rangle^{\;1}_{8} \big\rangle^{\;0}_{1}  \right) + 
\\ 
&\hspace{55ex} +\frac{1}{2} \left(\sqrt{\frac{2}{3}} \big| |c \bar{c} \rangle^{\;0}_{1}\otimes |c\bar{c}\rangle^{\;0}_{1} \big\rangle^{\;0}_{1}   +\sqrt{\frac{1}{3}} \big| |c \bar{c} \rangle^{\;0}_{8}\otimes |c\bar{c}\rangle^{\;0}_{8} \big\rangle^{\;0}_{1} \right),
\label{fierz_spin0_6x6}
\end{align}
\begin{align}
 \label{fierz_spin1} 
&\left| |cc\rangle_{\bar{3}}^{\;1} \otimes |\bar{c}\bar{c}\rangle_{3}^{\;1} \right\rangle_{1}^{\;1} = \sqrt{\frac{1}{3}} \big| |c \bar{c} \rangle^{\;0}_{1}\otimes |c\bar{c}\rangle^{\;1}_{1} \big\rangle^{\;1}_{1} -\sqrt{\frac{2}{3}} \big| |c \bar{c} \rangle^{\;0}_{8}\otimes |c\bar{c}\rangle^{\;1}_{8} \big\rangle^{\;1}_{1},\\
&\left| |cc\rangle_{\bar{3}}^{\;1} \otimes |\bar{c}\bar{c}\rangle_{3}^{\;1} \right\rangle_{1}^{\;2} = \sqrt{\frac{1}{3}} \big| |c \bar{c} \rangle^{\;1}_{1}\otimes |c\bar{c}\rangle^{\;1}_{1} \big\rangle^{\;2}_{1} -\sqrt{\frac{2}{3}} \big| |c \bar{c} \rangle^{\;1}_{8}\otimes |c\bar{c}\rangle^{\;1}_{8} \big\rangle^{\;2}_{1}.
 \label{fierz_spin2} 
\end{align}

The Fierz transformations of Eqs.~\eqref{fierz_spin0}---\eqref{fierz_spin2} are unitary transformations in spin and color space, which convert the orthonormal  $\left||cc\rangle \otimes |\bar{c}\bar{c}\rangle \right\rangle$  states to the orthonormal $\left||c\bar{c}\rangle \otimes |c\bar{c}\rangle \right\rangle$ states. In practice, these transformations act on  the spinor structure of the   ${\cal M}\left(gg\to \left[c\bar cc\bar c\right](J)+g\right)$ matrix elements. Since the initial state is a diquark-antidiquark configuration, the unpolarized cross-section calculations require summing over the spinor polarizations in the diquark-antidiquark basis. However with the help of these transformations ~\eqref{fierz_spin0}---\eqref{fierz_spin2}, this sum can be calculated in a more efficient way in the meson-like basis, for which the projection technique is well-established \cite{Guberina:1980dc,Kuhn:1979bb}.

Taking into account the coefficients in Eqs.~\eqref{fierz_spin0}---\eqref{fierz_spin2} and the fact that  $\beta^2 = 1 - \alpha^2$, we write down the structure of the total amplitude squared as a sum of the squared amplitudes for each of the $\left||c\bar c\rangle\otimes|c\bar c\rangle\right\rangle$ configurations:
\begin{align}\notag
&\big|{\cal A}^{(0)}\big|^2 = \sum_i \big|{\cal A}_i(gg\to T_{4c}(0^{++}) + g) \big|^2 = I_\textrm{ph.sp.}|\Psi^{(0)}(0,0,0)|^2\times 
\\
\label{main_spin0}
&\hspace{35ex}\times\bigg\lbrace \frac{6-5\alpha^2}{12}\big|{\cal B}_{11}^{11}\big|^2 + \frac{3-\alpha^2}{12}\big|{\cal B}^{11}_{88}\big|^2 + \frac{2+\alpha^2}{12}\big|{\cal B}^{00}_{11}\big|^2 + \frac{1+5\alpha^2}{12}\big|{\cal B}^{00}_{88}\big|^2 \bigg\rbrace,\\
&\big|{\cal A}^{(1)}\big|^2 = \sum_i \big|{\cal A}_i(gg\to T_{4c}(1^{+-})+g) \big|^2 = I_\textrm{ph.sp.}|\Psi^{(1)}(0,0,0)|^2\;\bigg\lbrace \frac{1}{3}\big|{\cal C}_{11}^{10}\big|^2 + \frac{2}{3}\big|{\cal C}^{10}_{88}\big|^2\bigg\rbrace, \\
&\big|{\cal A}^{(2)}\big|^2 = \sum_i \big|{\cal A}_i(gg\to T_{4c}(2^{++})+g) \big|^2 = I_\textrm{ph.sp.}|\Psi^{(2)}(0,0,0)|^2\;\bigg\lbrace \frac{1}{3}\big|{\cal D}_{11}^{11}\big|^2 + \frac{2}{3}\big|{\cal D}^{11}_{88}\big|^2\bigg\rbrace,
\label{main_spin2}
\end{align}
where $i$ runs over the  $\left||c\bar{c}\rangle \otimes |c\bar{c}\rangle \right\rangle$ configurations (see Eqs. \eqref{fierz_spin0}---\eqref{fierz_spin2} and below). The matrix elements $\{ {\cal B}_{11}^{11}~\ldots~{\cal D}^{11}_{88}\}$ are defined for the $T_{4c}(0^{++})$ as
\begin{align}
\label{comps_spin_0}
&{\cal B}_{11}^{11} \equiv {\cal M}\left(gg\to \big||c\bar c \rangle_1^{\;1}\otimes |c\bar c \rangle_1^{\;1}\,\rangle_1^{\;0} + g\right),
&{\cal B}_{88}^{11} \equiv {\cal M}\left(gg\to \big||c\bar c \rangle_8^{\;1}\otimes |c\bar c \rangle_8^{\;1}\,\rangle_1^{\;0} + g\right),\\ 
&{\cal B}_{11}^{00} \equiv {\cal M}\left(gg\to \big||c\bar c \rangle_1^{\;0}\otimes |c\bar c \rangle_1^{\;0}\,\rangle_1^{\;0} + g\right),
&{\cal B}_{88}^{00} \equiv {\cal M}\left(gg\to \big||c\bar c \rangle_8^{\;0}\otimes |c\bar c \rangle_8^{\;0}\,\rangle_1^{\;0} + g\right),
\end{align}
for the $T_{4c}(1^{+-})$ as
\begin{align}
\label{comps_spin_1}
&{\cal C}_{11}^{10} \equiv {\cal M}\left(gg\to \big||c\bar c \rangle_1^{\;1}\otimes |c\bar c \rangle_1^{\;0}\,\rangle_1^{\;1} + g\right),
&{\cal C}_{88}^{10} \equiv {\cal M}\left(gg\to \big||c\bar c \rangle_8^{\;1} \otimes |c\bar c \rangle_8^{\;0}\,\rangle_1^{\;1} + g\right),
\end{align}
and for the $T_{4c}(2^{++})$ as
\begin{align}
\label{comps_spin_2}
&{\cal D}_{11}^{11} \equiv {\cal M}\left(gg\to \big||c\bar c \rangle_1^{\;1}\otimes |c\bar c \rangle_1^{\;1}\,\rangle_1^{\;2} + g \right),
&{\cal D}_{88}^{11} \equiv {\cal M}\left(gg\to \big||c\bar c \rangle_8^{\;1}\otimes |c\bar c \rangle_8^{\;1}\,\rangle_1^{\;2} + g\right).
\end{align}
\indent 
We remark that by construction, the $\big||c\bar c\rangle\otimes |c\bar c\rangle\big\rangle$ configurations in the $\mathcal B$, $\mathcal C$, $\mathcal D$ components in Eqs.~\eqref{main_spin0}---\eqref{main_spin2} are orthogonal to each other both in color and spin spaces, and therefore  $|{\cal A}^{(J)}|^2$ --- the total amplitude squared  --- is free from the interference terms.

\section{Perturbative calculation}
\label{sec:calc}

\begin{figure}[t]
\includegraphics[width=1\linewidth]{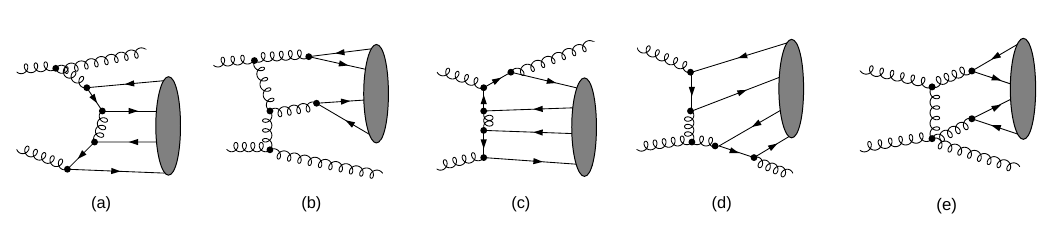}
\caption{Some of the Feynman diagrams for the $gg \to T_{4c}+g$ process: ISR (a-b), FSR (c-d), and Rest (e). The gray blob represents the tetraquark in the final state.}
\label{fig:diags}
\end{figure}

In the factorization scheme of Eq.~\eqref{main} we construct the matrix elements with the selected $\left||c\bar{c}\rangle \otimes |c\bar{c}\rangle \right\rangle_1^{}$ states, starting from the matrix elements with two heavy quark-antiquark pairs generated with ${\bf q_1}={\bf q_2}={\bf q_{12}}=0$. In these perturbative calculations, we work with the ${\cal O}(g_s^5)$ matrix elements for the $gg\to c\bar c c\bar c g$ partonic-level process, where all gluons, quarks and antiquarks are defined on the mass shell. We generate the full set of 682 tree matrix elements in Feynman gauge by means of the \texttt{FeynArts}~\cite{Hahn:2000kx} package. Figure~\ref{fig:diags} shows a diagrammatic sketch of the process.

Each matrix element contains a product of the two fermion lines due to  the two quark-antiquark pairs
\begin{equation}\label{vubarvubar}
\bar{u}^j(p_2,\lambda_2)\,{\cal O}_1^{ji}\,
 v^i(p_1,\lambda_1)\cdot
 \bar{u}^l(p_4,\lambda_4)\,
 {\cal O}_2^{lk}\,v^k(p_3,\lambda_3),
\end{equation}
where ${\cal O}_{1,2}$  denote some Dirac operators, which may depend on the external momenta, and each heavy (anti)quark carries one-quarter of the total tetraquark's momentum $P$: $p_1^2=p_2^2=p_3^2=p_4^2=\left(P/4\right)^2 = M^2/16$, with $M$ being the tetraquark mass.  In the present case, in which the two heavy quark-antiquark pairs have the same flavor, there are two spinor-antispinor couplings in the meson-like basis, $\lambda_1\lambda_2$-$\lambda_3\lambda_4$ and $\lambda_1\lambda_4$-$\lambda_2\lambda_3$, and both contribute to the $T_{4c}$ formation. We made sure that, after arranging  the two couplings  in the code, the 341 matrix elements matched the $\lambda_1\lambda_2$-$\lambda_3\lambda_4$ coupling and the other  341 matrix elements matched the $\lambda_1\lambda_4$-$\lambda_2\lambda_3$ coupling.

In order to construct the $\left||c\bar{c}\rangle \otimes |c\bar{c}\rangle \right\rangle_1$ states,
we need to select the two spinor-antispinor Fock states with the right spin and color. 
For the spin part, we make use of the standard quark-antiquark covariant projectors \cite{Guberina:1980dc,Kuhn:1979bb} in combination with the  appropriate color factors for the color singlet and  color octet:
\begin{eqnarray}
\label{proj}
     \Pi_\textrm{P}^{(a)}(p,m)=\frac{\slashed p- m}{2\sqrt{2}}\ \gamma^5 \otimes  {\cal C}^{(a)}\,,\qquad\qquad 
     \Pi_\textrm{V}^{\mu\,(a)}(p,m)=\frac{\slashed p- m}{2\sqrt{2}}\ \gamma^{\mu} \otimes  {\cal C}^{(a)}\,,\qquad\qquad
     {\cal C}^{(a)} = \left\lbrace\frac{\boldsymbol 1}{\sqrt{N_c}}, \frac{\lambda^a}{\sqrt{2}}\right\rbrace,
\end{eqnarray}
where P and V are used to denote the spin singlet and triplet, respectively,
$N_c=3$ is the number of colors  and $\lambda^a$ are the generators of the SU(3) color group.

We select the $\left||c\bar{c}\rangle \otimes |c\bar{c}\rangle \right\rangle_1$ states with the spin of $T_{4c}$ by summing over the spinor polarizations, $\lambda_1,\lambda_2,\lambda_3\,\lambda_4$,  and over the meson-like state polarization, $h_1,h_2$, at the level of the amplitude. The sum over the $\lambda_1,\lambda_2,\lambda_3\,\lambda_4$ and $h_1,h_2$ polarizations can be expressed through the spin part of the covariant projectors~\eqref{proj}, $\Pi_\textrm{P}(p,m)$ and $\Pi_\textrm{V}^{\nu}(p,m)$, 
as shown below.
\\
Spinor polarizaration sums for the case of $T_{4c}(0^{++})$:
\begin{align}
\label{pol01mm}
&\sum_{\lambda_1\,\lambda_2}\sum_{\lambda_3\,\lambda_4}\langle\frac{1}{2}\lambda_1;\frac{1}{2}\lambda_2 |0,0\rangle\langle\frac{1}{2}\lambda_3;\frac{1}{2}\lambda_4 |0,0\rangle\; v_{\lambda_1}^i \bar{u}_{\lambda_2}^j v_{\lambda_3}^k \bar{u}_{\lambda_4}^l = \left(\Pi_\textrm{P}(P/2,M/2)\right)^{ij}\:\left(\Pi_\textrm{P}(P/2,M/2)\right)^{kl},\\
\notag
&\sum_{h_1\,h_2}\sum_{\lambda_1\,\lambda_2}\sum_{\lambda_3\,\lambda_4}\langle 1,h_1; 1,h_2|0,0\rangle \langle\frac{1}{2}\lambda_1;\frac{1}{2}\lambda_2 |1,h_1\rangle \langle\frac{1}{2}\lambda_3;\frac{1}{2}\lambda_4 |1,h_2\rangle\; v_{\lambda_1}^i \bar{u}_{\lambda_2}^j v_{\lambda_3}^k \bar{u}_{\lambda_4}^l
= \\
&\hspace{62ex} =\left(\Pi_\textrm{V}^{\mu}(P/2,M/2)\right)^{ij}\,\left(\Pi_\textrm{V}^{\nu}(P/2,M/2)\right)^{kl}J_{\mu\nu}(P).
\label{pol02mm}
\end{align}  
Spinor polarization sum  for the case of $T_{4c}(1^{+-})$: 
\begin{multline}
\label{pol1mm}
\sum_{\lambda_1\,\lambda_2}\sum_{\lambda_3\,\lambda_4}\langle 1,h_1; 0,0|1,J_z\rangle \langle\frac{1}{2}\lambda_1;\frac{1}{2}\lambda_2 |1,h_1\rangle\langle\frac{1}{2}\lambda_3;\frac{1}{2}\lambda_4 |0,0\rangle\; v_{\lambda_1}^i \bar{u}_{\lambda_2}^j v_{\lambda_3}^k \bar{u}_{\lambda_4}^l = \\ = \left(\Pi_\textrm{P}(P/2,M/2)\right)^{ij}\:\left(\Pi_\textrm{V}^{\nu}(P/2,M/2)\right)^{kl}\varepsilon_{\nu}(P).
\end{multline} 
Spinor polarization sum for the case of $T_{4c}(2^{++})$: 
\begin{multline}\label{pol2mm}
\sum_{h_1\,h_2}\sum_{\lambda_1\,\lambda_2}\sum_{\lambda_3\,\lambda_4}\langle 1,h_1; 1,h_2|2,J_z\rangle \langle\frac{1}{2}\lambda_1;\frac{1}{2}\lambda_2 |1,h_1\rangle\langle\frac{1}{2}\lambda_3;\frac{1}{2}\lambda_4 |1,h_2\rangle\; v_{\lambda_1}^i \bar{u}_{\lambda_2}^j v_{\lambda_3}^k \bar{u}_{\lambda_4}^l = \\ = \left(\Pi_\textrm{V}^{\mu}(P/2,M/2)\right)^{ij}\:\left(\Pi_\textrm{V}^{\nu}(P/2,M/2)\right)^{kl}\varepsilon_{\mu\nu}(P).
\end{multline} 
For the sake of compactness, we give the spinor polarization sums for the $\lambda_1\lambda_2$-$\lambda_3\lambda_4$ coupling, while the expressions for the $\lambda_1\lambda_4$-$\lambda_2\lambda_3$ coupling can be retrieved under the permutation \{$\lambda_2\longleftrightarrow \lambda_4$, $j\longleftrightarrow l$\}. In the polarization sums above, we use the standard Clebsch-Gordan coefficients; $\Pi_\textrm{P}$ and $\Pi_\textrm{V}^{\mu}$ are the spin parts of the projectors~\eqref{proj};\; $J_{\mu\nu} = \frac{1}{\sqrt 3}(-g_{\mu\nu} + P_{\mu}P_{\nu}/M^2)$;  $\epsilon_{\mu}(P)$ and $\epsilon_{\mu\nu}(P)$ are the tetraquark spin-one polarization vector and spin-two polarization tensor, respectively. The polarization vector and tensor satisfy $\epsilon(P)\cdot P = 0~$ and $~\epsilon_{\mu\nu}(P)P^{\nu} = \epsilon_{\mu\nu}(P)P^{\mu} = 0$.

With the help of the analytic expressions of Eqs.~\eqref{pol01mm}---\eqref{pol2mm}, the two 
fermion lines, $\bar{u}_{\lambda_2}\:{\cal O}_1\:v_{\lambda_1}$ and $\bar{u}_{\lambda_4}\:{\cal O}_2\:v_{\lambda_3}$, are converted into traces. Then each of the matrix elements with the $\lambda_1\lambda_2$-$\lambda_3\lambda_4$ coupling acquires two traces, and each of the matrix elements with the $\lambda_1\lambda_4$-$\lambda_2\lambda_3$ coupling acquires one trace.
For example, in the case of $T_{4c}(0^{++})$, the application of Eq.~\eqref{pol01mm} gives
\begin{align}
\label{two-traces}
&\sum\limits_{\lambda_1\lambda_2}
\sum\limits_{\lambda_3\lambda_4}
\langle\frac{1}{2}\lambda_1;\frac{1}{2}\lambda_2 |0,0\rangle\langle\frac{1}{2}\lambda_3;\frac{1}{2}\lambda_4 |0,0\rangle
\;
\bar{u}_{\lambda_2}\,{\cal O}_1\,v_{\lambda_1}\,\bar{u}_{\lambda_4}\,{\cal O}_2\,v_{\lambda_3} =
\textrm{Tr}\,\Big\lbrace\Pi_{\text{P}}^{}\,{\cal O}_1\Big\rbrace\cdot\textrm{Tr}\Big\lbrace\Pi_{\text{P}}^{}\,{\cal O}_2 \Big\rbrace\,, \\
\label{one-trace}
&\sum\limits_{\lambda_1\lambda_4}
\sum\limits_{\lambda_2\lambda_3}
\langle\frac{1}{2}\lambda_1;\frac{1}{2}\lambda_4 |0,0\rangle\langle\frac{1}{2}\lambda_2;\frac{1}{2}\lambda_3|0,0\rangle
\;
\bar{u}_{\lambda_2}\,{\cal O}_1\,v_{\lambda_1}\,\bar{u}_{\lambda_4}\,{\cal O}_2\,v_{\lambda_3} = 
\textrm{Tr}\,\Big\lbrace\Pi_{\text{P}}^{}\,{\cal O}_1\,\Pi_{\text{P}}^{}\,{\cal O}_2\Big\rbrace 
\;.
\end{align}
This technique allows one to handle the spinors in the code analytically in a covariant way. The projection was carried out with our code in the framework of the \texttt{FeynCalc}~\cite{Shtabovenko:2020gxv} package. Further evaluation of the matrix elements containing traces was performed in~\texttt{FORM}~\cite{Kuipers:2012rf} computer algebra system. The implementation of the projection technique in our code was verified by reproducing the $g g \to J/\psi\, J/\psi$ analytic cross-section from \cite{Humpert:1983yj}.

After introducing the spinor polarization sums, it is convenient to address the phase space factor $I_\textrm{ph.sp.}$  in Eq.~\eqref{main}. For the phase space integration, one needs to be sure that the normalizations of the matrix element squared, wave function squared and invariant phase space measure are consistent with one another. It must be remembered that the polarization sums are written for spinors with the relativistic normalization $\bar{u}u = 2m_c$, and the $T_{4c}$ wave function satisfies the nonrelativistic normalization. One then needs to compensate only for the spinor normalization by a factor $I_{QQ}$ for each pair\,\footnote{It is fairly common to absorb $I_{QQ}$ into the projectors definition, but here we prefer to give it  in a factorized form.} and to compensate for the relativistic phase space  by a  factor $I_{4Q}$, such that the total correction factor in the colliding partos center-of-mass frame  is
\begin{equation}\label{norm}
I_\textrm{ph.sp.} = I_{4Q} \times I_{QQ}\times I_{QQ} 
= 4\sqrt{2}\,M^{-3/2}, 
\end{equation}
where $I_{4Q}=\sqrt{2 M}$ and  $I_{QQ} = 1/m_{QQ} = 2/M$.

Now let us address the diagrams that represent the $gg\to T_{4c} + g$ scattering matrix elements. The full set of diagrams can be regarded as the set of three groups: initial state radiation (ISR), where a gluon is emitted from an incoming gluon line, final state radiation (FSR), where a gluon is emitted from an outgoing quark line; and what we call the rest (Rest), where a gluon is emitted from an internal line (either a quark or a gluon line) or a four-gluon vertex. As an example, in Fig.~\ref{fig:diags} we display two ISR diagrams~(a,b), two FSR diagrams~(c,d), and one Rest diagram~(e). We remark that the diagrams with a four-gluon vertex occur in each of the three groups.

The abovementioned groups contribute differently to $C$-even and $C$-odd tetraquark production. For example,   the ISR group~(144 diagrams) contributes only to the amplitudes with a $C$-even tetraquark in the final state. This can be explained by $C$-parity conservation in the $gg \to |cc\bar{c}\bar{c}(J^{PC})\rangle_1$ reaction, which prohibits the production of $T_{4c}(1^{+-})$ at threshold in the gluon fusion channel. We checked that this prohibition remains valid when one of the two gluons goes off-shell, which is exactly what occurs in the ISR process.

Another type of selection rule concerns the case of two-trace projection onto the color-singlet $|c\bar c \rangle$ structures. If $C$-parity is conserved in the $g^{(*)}\ldots g^{(*)}\to |c\bar c \rangle_1$ reaction, with the gluons allowed to be off-shell, then a nonzero contribution can arise only from the diagrams in which any fermion line associated with a vector $|c\bar{c}\rangle_1$ structure is coupled to an odd number of gluons~(three), whereas any fermion line associated with a pseudoscalar $|c\bar{c}\rangle_1$ structure is coupled to an even number of gluons~(two or four). According to our calculations, this selection rule proves to be valid, i.e. for the production process studied, $C$-parity is indeed conserved in the abovementioned reaction. At the same time, as expected, we saw that this selection rule does not hold in the $g^{(*)}\ldots g^{(*)}\to |c\bar c \rangle_8$ reaction.

As a result of our calculations, we ascertain that the matrix elements for the $C$-even tetraquarks with $J^{PC}=\{0^{++},2^{++}\}$ are proportional to the fully anti-symmetric color structure $f_{ABC}$. Conversely, the matrix elements for the $J^{PC}=1^{+-}$ tetraquark contain both the fully symmetric and fully anti-symmetric color structures $d_{ABC}$ and $f_{ABC}$, however only the part that is proportional to $d_{ABC}$ survives in the squared matrix element. In other words, we found a correlation between the color structure and the $C$-parity of the $T_{4c}+g$ final state: $f_{ABC}$ in the case of negative final state $C$-parity, and  $d_{ABC}$ in the case of positive final state $C$-parity. It is worth recalling that the same correlation between the final state $C$-parity and color structure has been found in the background process of gluon fusion to double charmonia with a real gluon emission~\cite{Likhoded:2016zmk}.

After taking the Dirac traces (see Eqs. \eqref{two-traces}$-$\eqref{one-trace}) and calculating the color factors, the matrix elements can be written down formally with all spatial and color indices given explicitly. We use the Lorentz indices $\{\mu_1,\,\mu_2,\,\mu_3\}$ and the indices $\{a_1,\,a_2,\,a_3\}$ for the adjoint representation of $SU(3)$ color for the two initial and one final gluons with the polarization vectors $\{\epsilon_{\mu_1}(k_1),\epsilon_{\mu_2}(k_2),\epsilon^{\ast}_{\mu_3}(k_3)\}$, respectively. The eight matrix elements introduced in Section~\ref{sec:phenomen} can be written as
\begin{align}
\label{comps_spin_0_struct}
&{\cal B}_{11}^{11}\Big|_{a_{1\,}a_{2\,}a_{3}} = f_{a_{1\,}a_{2\,}a_{3}}\:{\cal M}^{\mu_1\,\mu_2\,\mu_3\,\mu\,\nu}_\textrm{VV}(k_1,k_2,k_3,P)\; J_{\mu\nu}(P)\,\epsilon_{\mu_1}\,\epsilon_{\mu_2}\,\epsilon^{\ast}_{\mu_3},\\
&{\cal B}_{88}^{11}\Big|_{a_{1\,}a_{2\,}a_{3}} = f_{a_{1\,}a_{2\,}a_{3}}\:{\cal M}^{\mu_1\,\mu_2\,\mu_3\,\mu\,\nu;\; a\, b}_\textrm{VV}(k_1,k_2,k_3,P)\; \frac{\delta_{a b}}{\sqrt{8}}\;J_{\mu\nu}(P)\,\epsilon_{\mu_1}\,\epsilon_{\mu_2}\,\epsilon^{\ast}_{\mu_3},\\ 
&{\cal B}_{11}^{00}\Big|_{a_{1\,}a_{2\,}a_{3}} = f_{a_{1\,}a_{2\,}a_{3}}\:{\cal M}_\textrm{PP}^{\mu_1\,\mu_2\,\mu_3}(k_1,k_2,k_3,P)\;\epsilon_{\mu_1}\,\epsilon_{\mu_2}\,\epsilon^{\ast}_{\mu_3},\\
&{\cal B}_{88}^{00}\Big|_{a_{1\,}a_{2\,}a_{3}} = f_{a_{1\,}a_{2\,}a_{3}}\:{\cal M}_\textrm{PP}^{\mu_1\,\mu_2\,\mu_3;\; a\, b}(k_1,k_2,k_3,P)\;\frac{\delta_{a b}}{\sqrt{8}}\,\epsilon_{\mu_1}\,\epsilon_{\mu_2}\,\epsilon^{\ast}_{\mu_3},\\
&{\cal C}_{11}^{10}\Big|_{a_{1\,}a_{2\,}a_{3}} = d_{a_{1\,}a_{2\,}a_{3\,}}{\cal M}^{\mu_1\,\mu_2\,\mu_3\,\mu}_\textrm{VP}(k_1,k_2,k_3,P)\;\epsilon_{\mu}(P)\,\epsilon_{\mu_1}\,\epsilon_{\mu_2}\,\epsilon^{\ast}_{\mu_3},\\
&{\cal C}_{88}^{10}\Big|_{a_{1\,}a_{2\,}a_{3}} = d_{a_{1\,}a_{2\,}a_{3\,}}{\cal M}^{\mu_1\,\mu_2\,\mu_3\,\mu;\; a\, b}_\textrm{VP}(k_1,k_2,k_3,P)\;\frac{\delta_{a b}}{\sqrt{8}}\,\epsilon_{\mu}(P)\,\epsilon_{\mu_1}\,\epsilon_{\mu_2}\,\epsilon^{\ast}_{\mu_3},\\
\label{comps_spin_1_struct}
&{\cal D}_{11}^{11}\Big|_{a_{1\,}a_{2\,}a_{3}} = f_{a_{1\,}a_{2\,}a_{3}}\:{\cal M}^{\mu_1\,\mu_2\,\mu_3\,\mu\,\nu}_\textrm{VV}(k_1,k_2,k_3,P)\;\epsilon_{\mu\nu}(P)\,\epsilon_{\mu_1}\,\epsilon_{\mu_2}\,\epsilon^{\ast}_{\mu_3},\\
&{\cal D}_{88}^{11}\Big|_{a_{1\,}a_{2\,}a_{3}} = f_{a_{1\,}a_{2\,}a_{3}}\:{\cal M}^{\mu_1\,\mu_2\,\mu_3\,\mu\,\nu;\; a\, b}_\textrm{VV}(k_1,k_2,k_3,P)\;\frac{\delta_{a b}}{\sqrt{8}}\,\epsilon_{\mu\nu}(P)\,\epsilon_{\mu_1}\,\epsilon_{\mu_2}\,\epsilon^{\ast}_{\mu_3},
\label{comps_spin_2_struct}
\end{align}
where the lowercase symbols $\{\textrm{VV},\textrm{PP},\textrm{VP}\}$ refer to  the polarization sums \eqref{pol01mm}---\eqref{pol2mm}. The factor $\left(\delta_{a b}/\sqrt{8}\right)$ ensures the correct normalization for the octet-octet components obtained with the color factor ${\cal C}^a = \lambda^a/\sqrt{2}$\; in the projectors~\eqref{proj}.

We analytically square the matrix elements  in~\texttt{FORM} and apply the completeness relations in order to sum over all spin states. Since, in collinear factorization, one assumes on shell partons in the initial and final states of the scattering process,  all three gluons must carry the transverse polarizations only. To get rid of the unwanted components in the gluon polarization vectors, we select  the physical states with the help of the completeness relation in the axial gauge~\cite{Baier:1983va,Humpert:1983yj}
\begin{equation}\label{gluon_sum}
\sum_{\lambda} \epsilon_{\alpha'}^{*}(k,\lambda) \epsilon_{\alpha}(k,\lambda) \Longrightarrow -g_{\alpha' \alpha} +\frac{k_{\alpha'}n_{\alpha}+k_{\alpha}n_{\alpha'}}{k \cdot n}- n^2 \frac{k_{\alpha'} k_{\alpha}}{(n \cdot k)^2},
\end{equation}
where $n$ is an auxiliary vector satisfying $n \cdot k \ne 0$. We choose $n = (1,0,0,0)$ for all the three gluon polarization sums. The sum over the polarizations of spin-one and spin-two $T_{4c}$ is carried out similarly through the completeness relations for the massive boson polarizations. The gauge invariance of each amplitude in Eqs.~\eqref{comps_spin_0_struct}--\eqref{comps_spin_2_struct} was verified analytically.

Under the approximation used in Eq.~\eqref{main}, the cross-section can be expressed as the product of the wave function squared and the short-distance factor, which encodes the information on the hard process. Since there is some freedom in the definition of the short-distance factor,  we introduce it as a dimensionless factor $F^{(J)}$ matched to the reduced wave function $\phi^{(J)} = |\Psi^{(J)}(0,0,0)|^2/M^9$. We also replace the  Mandelstam variables with the dimensionless ones: $\hat{s} = s M^2$, $\hat{t} = t M^2$, and $\hat{u} = u M^2=(1-s-t) M^2$. Thus, for each $J$, the modulus squared of the production amplitude summed  
 over the spin and color of the initial and final  states and multiplied by the spin-color averaging factor  can be written as 
\begin{equation}\label{amp_sqr}
\frac{1}{64\cdot 4}\sum\limits_{\textrm{col, spin}}\big|{\cal A}^{(J)}\big|^2 = ~\phi^{(J)}\times  (4\pi\alpha_s)^5\times F^{(J)}(s,t),
\end{equation}
where $F^{(J)}(s,t)$ is a linear combination of a few components,
\begin{align}\label{sdcs}
&F^{(0)}(s,t) =  \frac{6-5\alpha^2}{12} F_{11}^{11}(s,t) + \frac{3-\alpha^2}{12}F_{88}^{11}(s,t) + \frac{2+\alpha^2}{12}F_{11}^{00}(s,t) + \frac{1+5\alpha^2}{12}F_{88}^{00}(s,t),\\
&F^{(1)}(s,t) = \frac{1}{3} F_{11}^{10}(s,t) + \frac{2}{3} F_{88}^{10}(s,t),\\
&F^{(2)}(s,t) = \frac{1}{3} \widetilde{F}_{11}^{11}(s,t) + \frac{2}{3} \widetilde{F}_{88}^{11}(s,t).
\end{align}
For each component of $F^{(J)}$, we obtain the analytic expression in the form of a rational polynomial in the variables $s$, $t$. Our calculations also showed that $F_{88}^{10}(s,t) = 2F_{11}^{10}(s,t)$ and $\widetilde{F}_{88}^{11}(s,t) = 2\widetilde{F}_{11}^{11}(s,t)$.

After splitting the full set of diagrams into ISR, FSR and Rest groups in the code, we 
managed to ascertain that only the ISR  diagrams generate a divergence in the total amplitude. Whereas the contribution of the FSR+Rest groups alone is IR safe, the ISR contribution is responsible for a divergent behavior in the
soft and collinear limits.\,\footnote{One can disentangle the IR divergences by splitting the phase space into three regions that lead to the following three contributions to the total cross-section: (1) finite cross-section, (2) cross-section in the soft ($s\to 1$) and soft-collinear ($s\to 1$ and either $t \to 0$ or $u \to 0$) limits and (3) the cross-section in the hard-collinear ($t\to 0$ or $u\to 0$) limit. The soft and soft-collinear limits correspond to the production at threshold, $\sigma_{2\to 1} \sim \delta(p_T)$. Since in this article we consider a real gluon emission, we are not interested in the soft limit and we address only the correction for the hard-collinear limit.} As mentioned earlier, the ISR contribution vanishes in the case of $1^{+-}$ tetraquark production. Therefore, we  find the $F^{(1)}(s,t)$ expression to be finite in the $t\to (1-s)$ and $t\to 0$ limits. Meanwhile, the $F^{(0)}(s,t)$ and $F^{(2)}(s,t)$ expressions diverge due to the $ t\,(1-s-t)$ factor in their denominators.  Thus, $\widehat\sigma(gg\to T_{4c}(1^{+-})+g)$ is the only partonic cross-section which is collinear stable by default. The collinear stable expressions for $\widehat\sigma(gg\to T_{4c}(0^{++})+g)$ and $\widehat\sigma(gg\to T_{4c}(2^{++})+g)$ can be obtained by considering an evolution of the parton distribution functions (PDFs) via the DGLAP equations. With this technique the $1/\varepsilon_\textrm{IR}$ poles, having been isolated after performing the phase space integration with dimensional regularization, can be absorbed into the evolving PDFs.

In this article, we propose a simpler solution: applying a $p_T$ cut-off consistent with an experimental low $p_T$ cut. Hence, we expect that, starting from some value in the transverse momentum, our calculated cross-sections reproduce the cross-sections with the correct behavior
in the low $p_T$ limit. Our choices of $p_T$ cuts for the LHC experiments are discussed in Subsection~\ref{subsec:cuts}.

\section{cross-sections in the kinematic regions of interest for the LHC experiments}
\label{sec:res}
\subsection{Kinematic regions}
\label{subsec:cuts}
So far, observations of $T_{4c}$ candidates have been made  in the $J/\psi \left(\to \mu^+\mu^-\right) J/\psi \left(\to \mu^+\mu^-\right)$ and $J/\psi \left(\to \mu^+\mu^-\right) \psi(2S) \left(\to \mu^+\mu^-\right)$  decay modes. However, in the experimental articles, the cuts are often reported for individual muons or di-muons, rather than for the entire di-$J/\psi$ system (hereafter, we refer to the di-$J/\psi$ mode, but the same applies to the $J/\psi\,\psi(2S)$ mode in the case of ATLAS). In the LHCb article~\cite{LHCb:2020bwg}, it is explicitly stated that, in the $p_T$-threshold approach, the di-$J/\psi$ candidates were selected with the requirement $p_T^{\text{di-}J/\psi} > 5.2~\text{GeV}$. By contrast, ATLAS and CMS experiments report only selection cuts on $p_T(\mu)$ or $p_T(\mu^+\mu^-)$ in~\cite{ATLAS:2023bft,CMS:2023owd}. In ATLAS, two of the four selected muons must have $p_T > 4~\text{GeV}$, and the remaining two must have $p_T > 3~\text{GeV}$. In CMS, the low $p_T$ cut on each of the di-muons is $p_T(\mu^+\mu^-) > 3.5~\text{GeV}$. Each experiment specifies an acceptance range in terms of the pseudorapidity of individual muons: $|\eta(\mu)| < 2.5$ in ATLAS; $|\eta(\mu)| < 2.4$ in CMS, and $2 < \eta(\mu) < 5$ in LHCb. Meanwhile, none of the experiments applies the cuts on the di-$J/\psi$ pseudorapidity.

Since our code is designed solely for the production process, we do not track the subsequent decay of the produced tetraquark and we are therefore unable to impose cuts on its decay products. Therefore, we impose cuts on the phase space volume expressed in the tetraquark's transverse momentum $p_T$ and rapidity $y$ variables. We consider prompt production in the three kinematic regions of interest for the LHC experiments: the forward kinematic region $2 < y < 5,\quad 5 < p_T(\text{GeV}) < 50$  and the central kinematic region, $|y| < 2$, with the two $p_T$ ranges, $10 < p_T(\text{GeV}) < 100$ and  $20 < p_T(\text{GeV}) < 100$. In the case of the forward kinematic region, we assume that our cuts approximately correspond to the LHCb ones.
The cases of ATLAS and CMS are more complicated since there are, as yet, no measurements with given fiducial cuts on the di-$J/\psi$ transverse momentum and pseudorapidity. Even though the signal events can be collected at lower $p_T$, $i.e.$ even below 7 GeV for CMS and 14 GeV for  ATLAS, we recognize that the efficiency is poorly defined near these thresholds.
For this reason, we prefer to use more conservative $ p_T$ ranges that are accessible to ATLAS and CMS not only for the yield measurements but also for the cross-section measurements.
In the future, we can readily reproduce our calculations and refine the results according to the known fiducial cuts in terms of $p_T^{\text{di-}J/\psi} $  and $y^{\text{di-}J/\psi} $ variables.

\subsection{Hadronic cross-sections}
\label{subsec:sigma}
We calculate the hadronic cross-sections in the framework of the collinear factorization of the initial partons. As mentioned earlier, the partonic cross-section in the gluon-gluon fusion channel, convoluted with the gluon density functions $G(x_{1,2},\mu)$, constitutes the dominant part of the cross-section at LHC. For the numerical computations, we make use of the Monte Carlo integration coded with a \texttt{C++} macro linked to the \texttt{ROOT}~\cite{Brun:1997pa} environment. Since we deal with a two-body phase space, we find it more convenient to use private code rather than applying built-in algorithms for the phase space integration.
The  proton-proton cross-section reads 
\begin{equation}\label{sig_hadron}
\sigma_{pp} = \iint\, dx_1 dx_2\, G(x_1,\mu)\, G(x_2,\mu) \int\limits dt\, \frac{d\widehat{\sigma}(s,t,\mu)}{dt},
\end{equation}
where, for the gluon density functions $G(x_{1,2},\mu)$ and the strong coupling $\alpha_s(\mu)$, we apply the CTEQ18 NLO parametrization~\cite{Hou:2019efy}. We do not separate the renormalization scale in $\alpha_s(\mu)$ and the factorization scale in  $G(x_{1,2},\mu)$. For the scale range, we follow the conventional choice  $E_T/2 \leq \mu \leq 2 E_T$, with $E_T=\sqrt{p_T^2 + M^2}$ being the transverse energy.

The spectroscopy inputs, namely the masses, the four-body wave functions at the origin, and the mixing parameter of the color configurations in $0^{(\prime)++}$ states, are drawn from the nonrelativistic quark model calculations~\cite{Wang:2021kfv} and~\cite{Lu:2020cns}. In Appendix~\ref{app:spectr_inputs}, we list the spectroscopy inputs for $1S$, $2S$ and $3S$ states of the $T_{4c}$ tetraquark with quantum numbers $J^{PC}=\{0^{++},1^{+-},2^{++}\}$.
In Appendix~\ref{app:spectr_inputs} we also include  the root mean square radii used to extract the wave functions at the origin.

In the CSM, the $J^{PC}$ dependence of the cross-section distribution shapes comes from the short-distance factor, whereas the wave function magnitude affects only the normalization. Note also that, for each $J^{PC}$, the obtained expression of the short-distance factor $F^{(J)}$ holds for any radial excitation $nS$. Therefore, it is reasonable to factor out a radial state-dependent dimensionless factor $\Phi^{(J)}(nS)$ as follows
\begin{eqnarray}\label{Phi_factor}
    \sigma = \Phi(nS)\:\overline{\sigma}(\mu,\langle M\rangle,\langle\alpha^2\rangle),\qquad\qquad \Phi(nS) = 10^{10}\phi(nS)~~(\textrm{as given in Appendix~\ref{app:spectr_inputs}}),
\end{eqnarray}
where the $J$ notation is omitted, $\phi$ is the reduced wave function introduced in Eq.~\eqref{amp_sqr}, and $\langle M\rangle$ and $\langle\alpha^2\rangle$ are the mean values of $M$ and $\alpha^2$ obtained from averaging over the radial state masses. As we discuss later in the text, the impact of varying the $M$ and $\alpha^2$ numerical inputs is small compared to the scale uncertainty. In Eq.~\eqref{Phi_factor} we refer to $\sigma$ and $\overline{\sigma}$ as the full and reduced cross-sections, respectively.

The reduced cross-section $\overline{\sigma}$ represents the universal result of the pQCD calculation, which can be matched to the tetraquark wave functions obtained by solving
the four-body problem in any model. As yet, there is no solid understanding of how to assign the quantum numbers to the experimentally observed states. The experimentally observed states might not be the ground states, but the radial excitations with higher predicted masses or the superpositions of two or more unresolved neighboring states. Moreover, for some states, the four-body wave functions, predicted, for example, in~\cite{Wang:2021kfv} and in~\cite{Lu:2020cns}, differ by an order of magnitude~(see Appendix~\ref{app:spectr_inputs}). For these reasons, we prefer to present the cross-sections in the form of the reduced cross-sections $\overline{\sigma}$.

In the full cross-section $\sigma$ there are two main sources of uncertainties: the uncertainty in the wave function magnitude and the uncertainty in the scale definition. Their separation in  Eq.~\eqref{Phi_factor}  does take place, since the $\overline{\sigma}$ dependence on $M$ and $\alpha^2$ is small compared to the dependence on scale $\mu$ (at least at the leading-order accuracy). We saw that, for the $0^{++}$ or $2^{++}$ states, a mass variation within the range of experimental measurements \cite{LHCb:2020bwg,ATLAS:2023bft,CMS:2023owd} results in at most a 10\% deviation in $\overline{\sigma}(\langle M\rangle)$ calculated with the average over all the measured masses, $\langle M\rangle \simeq 6900 $ MeV. The same 10\%  deviation limit holds  
for the $1^{+-}$ state cross-section calculated with  $\langle M\rangle \simeq 6800 $ MeV, which is the average over all the predicted $1^{+-}$ masses~\cite{Lu:2020cns,Wang:2021kfv}.
In the case of the $0^{++}$ state, the reduced cross-section $\overline{\sigma}$ remains dependent on the weight parameter $\alpha^2$. The variation of $\alpha^2$ from its mean value $\langle\alpha^2\rangle = 0.5$ to $\alpha^2 = 0$ (corresponding to the pure $[6\otimes\bar{6}]$ color configuration) or $\alpha^2 = 1$ (corresponding to the pure $[\bar{3}\otimes 3]$ color configuration) does not significantly alter the magnitude of $\overline{\sigma}$ — the change being within  10\%. Meanwhile, the scale prescription remains the primary source of the $\overline{\sigma}$ uncertainty and reaches up to 50\% for the conventional choice mentioned above. This significant scale dependence is partially caused by the fact that $\widehat{\sigma} \sim \alpha_s^5(\mu)$.

\begin{figure}[t]
\centering
\begin{subfigure}[h]{0.49\linewidth}
\includegraphics[width=1\linewidth]{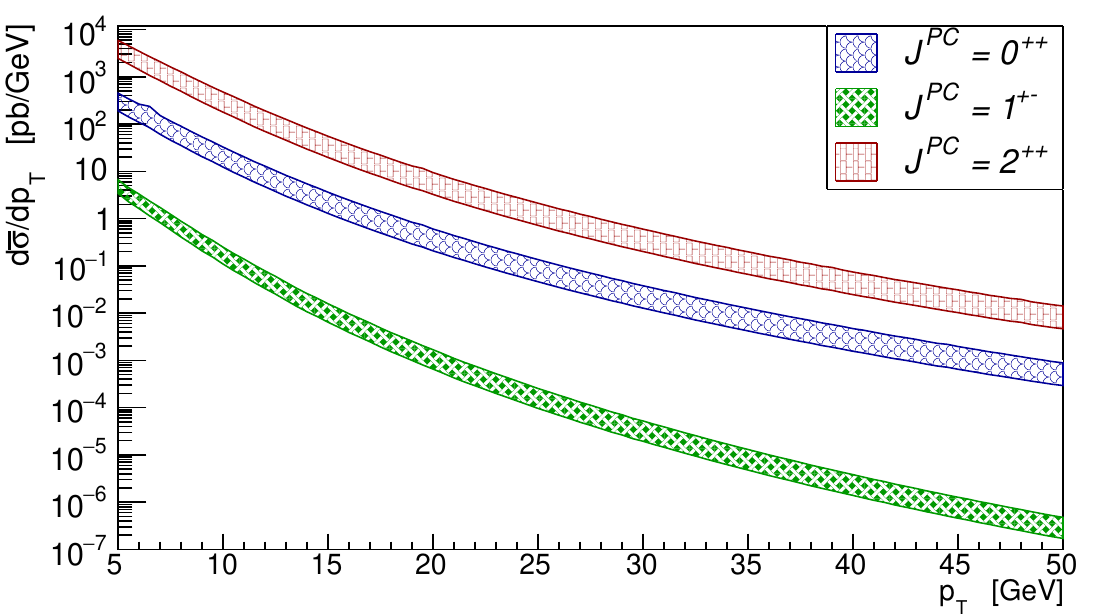}
\caption{The band corresponds to the $E_T/2 \leq \mu \leq 2 E_T$ scale variation, while $M$ and $\alpha^2$ are fixed to their average values.}
\label{fig:pt-distrib-forward-a}
\end{subfigure}
\hfill
\begin{subfigure}[h]{0.49\linewidth}
\includegraphics[width=1\linewidth]{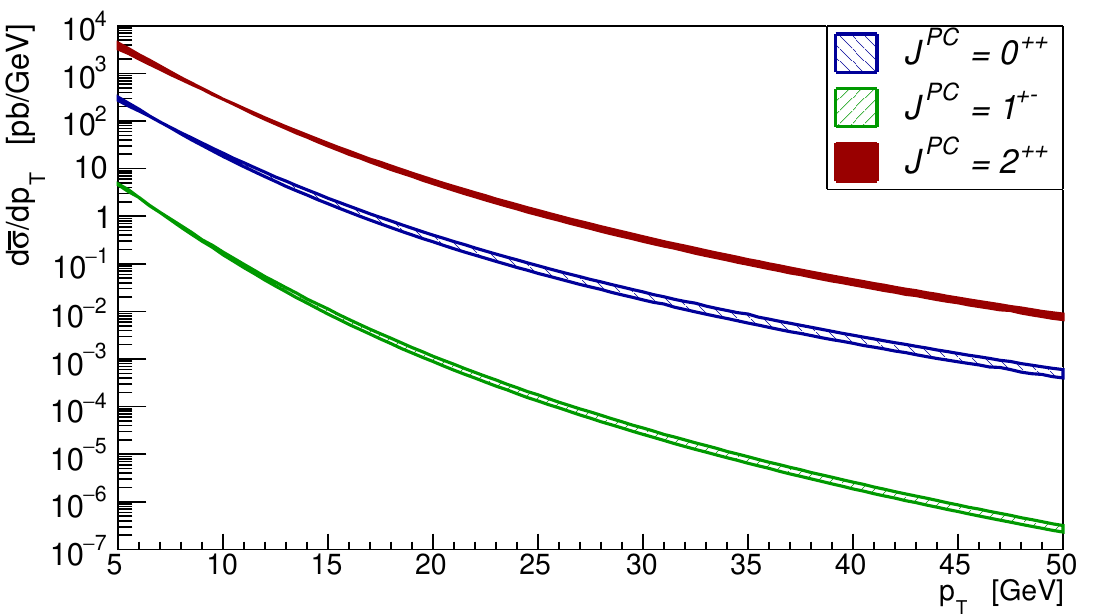}
\caption{The band corresponds to the variation of 
 the spectroscopy parameters $M$ and $\alpha^2$, while the scale is fixed to $E_T$. }
\label{fig:pt-distrib-forward-b}
\end{subfigure}
\caption{$d\overline{\sigma}/dp_T$ as a function of $p_T$ for the $pp\to T_{4c}(J^{PC})+X$ process within the forward acceptance $2 < y < 5$ and $p_T$-window $5 < p_T(\textrm{GeV}) < 50$. 
}
\label{fig:pt-distrib-forward}
\vspace{2ex}
\begin{subfigure}[h]{0.49\linewidth}
\includegraphics[width=1\linewidth]{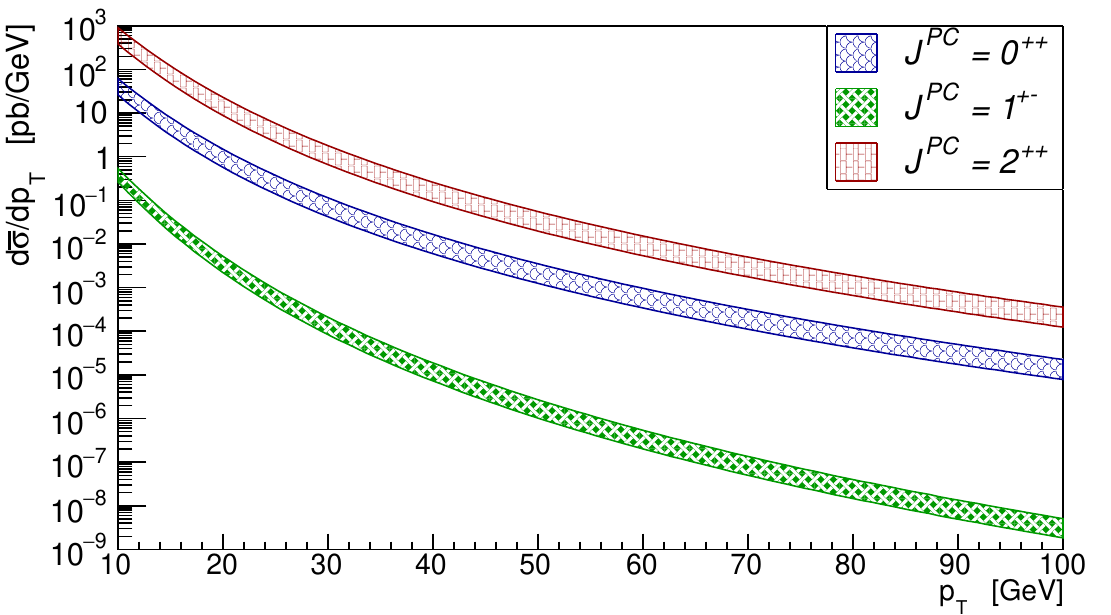}
\caption{The band corresponds to the $E_T/2 \leq \mu \leq 2 E_T$ scale variation, while $M$ and $\alpha^2$ are fixed to their average values.}
\label{fig:pt-distrib-central-a}
\end{subfigure}
\hfill
\begin{subfigure}[h]{0.49\linewidth}
\includegraphics[width=1\linewidth]{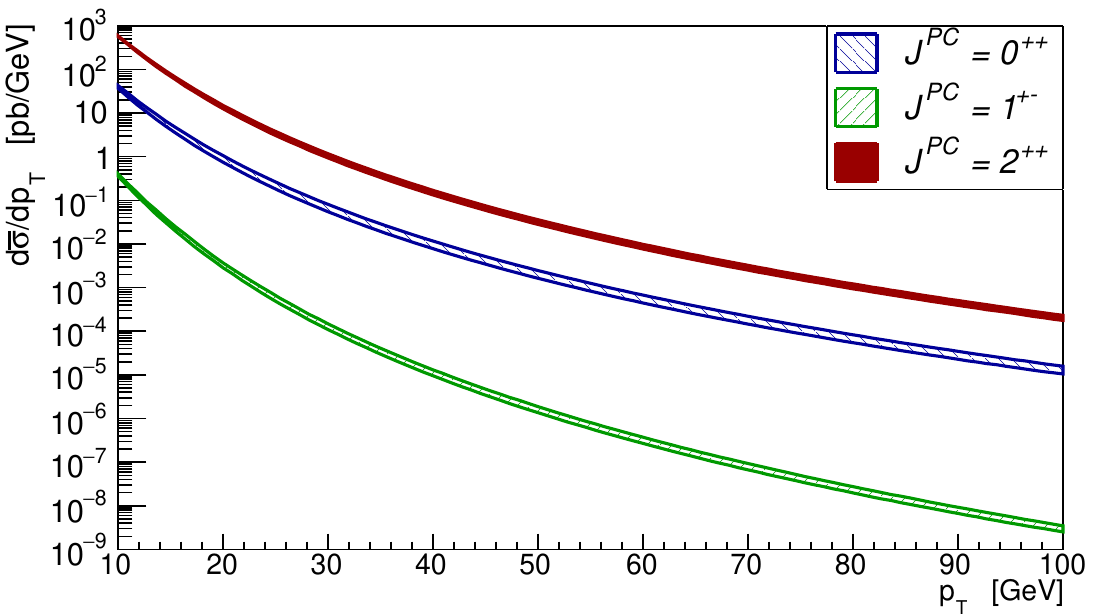}
\caption{The band corresponds to the variation of 
 the spectroscopy parameters $M$ and $\alpha^2$, while the scale is fixed to $E_T$.}
\label{fig:pt-distrib-central-b}
\end{subfigure}
\caption{$d\overline{\sigma}/dp_T$ as a function of $p_T$ for the $pp\to T_{4c}(J^{PC})+X$ process within the central acceptance $|y| < 2$ and  $p_T$-window $10 < p_T(\textrm{GeV}) < 100$.}
\label{fig:pt-distrib-central}
\end{figure}

The differential cross-sections were calculated straightforwardly by filling out the ROOT histograms when running through the Monte Carlo loop in the triple integration~\eqref{sig_hadron}. Figures~\ref{fig:pt-distrib-forward},~\ref{fig:pt-distrib-central} present the plots for the $p_T$-spectra at the collision energy $\sqrt{S} = 13~\text{TeV}$. Subfigures~\ref{fig:pt-distrib-forward-a},~\ref{fig:pt-distrib-central-a}  show $d\overline{\sigma}/dp_T$, calculated with  the approximate mean mass inputs  $\langle M(2^{++}) \rangle  = \langle M(0^{++}) \rangle  \simeq  6900~\text{MeV}$  
and $\langle  M(1^{+-}) \rangle \simeq  6800~\text{MeV}$, where  each $d\overline{\sigma}/dp_T$ distribution is plotted with a band corresponding to $E_T/2 \leq \mu \leq 2E_T$. 
Subfigures~\ref{fig:pt-distrib-forward-b},~\ref{fig:pt-distrib-central-b} show $d\bar{\sigma}/dp_T$, calculated at the central scale $\mu=E_T$, where the band corresponds to the spectroscopy input variation: the  $M(0^{++})$ and $M(2^{++})$ masses run in the 6400 MeV $\div$ 7300 MeV range (the approximate lowest and highest values from~\cite{LHCb:2020bwg,ATLAS:2023bft,CMS:2023owd}); the  $M(1^{+-})$ mass runs in the  6480 MeV $\div$ 7050 MeV range (the approximate lowest and highest values from~\cite{Lu:2020cns,Wang:2021kfv}); and $0\leq \alpha^2 \leq 1$ (all possible values).

In Figures~\ref{fig:pt-distrib-forward},~\ref{fig:pt-distrib-central}, we can notice that the dependence on the spectroscopy parameters is much less significant than the scale dependence. The production cross-section for the $1^{+-}$ state is suppressed by $2$ to $3$ orders of magnitude compared to $C$-even states, and exhibits a faster decrease with $p_T$.  The differential cross-sections $d\overline{\sigma}/dp_T$ for  $0^{++}$ and $2^{++}$ tetraquarks scale as $1/p_T^6$, while for the $1^{+-}$ tetraquark as $1/p_T^8$. 
Another important result that is immediately seen in Figures~\ref{fig:pt-distrib-forward}, \ref{fig:pt-distrib-central} is the tenfold enhancement in $\overline{\sigma}(2^{++})$ compared to $\overline{\sigma}(0^{++})$.

\begin{figure}[t]
\centering
\begin{subfigure}[h]{0.48\linewidth}
\includegraphics[width=1\linewidth]{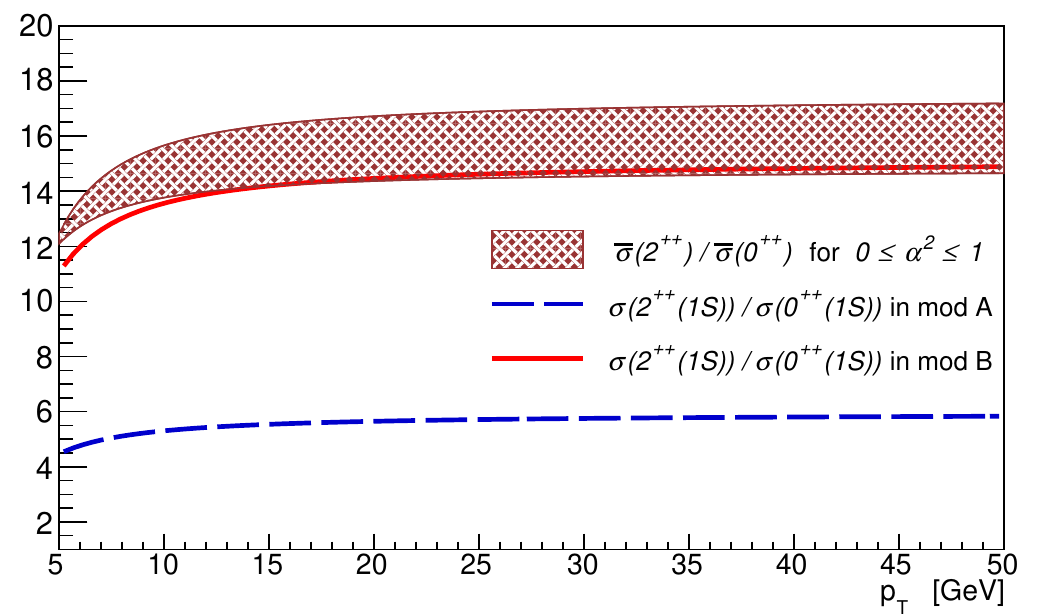}
\caption{Acceptance $2 < y < 5$.}
\end{subfigure}
\hfill
\begin{subfigure}[h]{0.48\linewidth}
\includegraphics[width=1\linewidth]{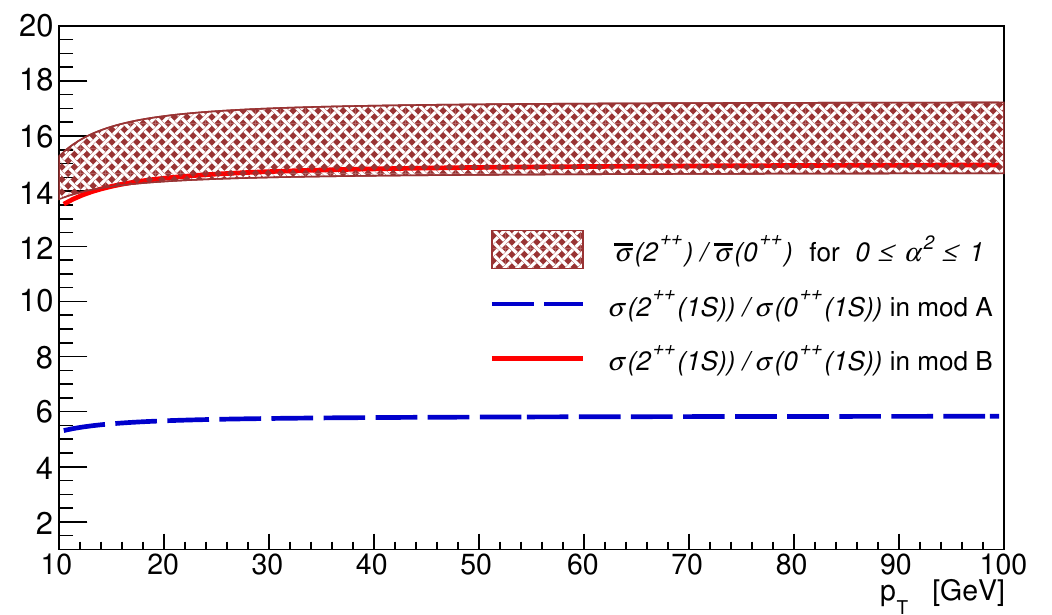}
\caption{Acceptance $|y| < 2$.}
\end{subfigure}
\caption{$\overline{\sigma}(pp\to T_{4c}(2^{++})+X)/\overline{\sigma}(pp\to T_{4c}(0^{++})+X)$ and $\sigma(pp\to T_{4c}(2^{++}(1S))+X)/\sigma(pp\to T_{4c}(0^{++}(1S))+X)$ ratios 
as functions of $p_T$ in the forward (a) and central (b) acceptance ranges.   Model A refers to~\cite{Wang:2021kfv}, and model B refers to~\cite{Lu:2020cns}.
}
\label{fig:pt-ratios}
\end{figure}

Figure~\ref{fig:pt-ratios} shows the ratio between the production cross-sections of the $2^{++}$ and $0^{++}$ tetraquark states,  in the central and forward kinematic regions.
The scale dependence in this ratio cancels out, making it possible to isolate the dependence on $\alpha^2$. By analogy with the reduced cross-section, we define the reduced ratio as $r = \overline{\sigma}(pp\to T_{4c}(2^{++})+X)/\overline{\sigma}(pp\to T_{4c}(0^{++})+X)$ evaluated at the mean mass $\langle M \rangle \simeq 6900$~MeV. Our results for $r$ are shown in Figure~\ref{fig:pt-ratios} as a function of $p_T$ with a band between the upper curve, corresponding to $\alpha^2=0$, and the lower curve, corresponding to $\alpha^2=1$. As one can see, the reduced ratio does not depend on the acceptance range and it slightly rises with $p_T$, saturating to a limit value. We observe that, for $5 < p_T (\text{GeV}) < 100$, the interval is  $12 \le r(p_T) \le 17$. This interval can be readily converted to an interval for the $2^{++}(nS)$ to  $0^{++}(nS)$ cross-section ratio by multiplying by  $\phi^{(2)}(nS)/\phi^{(0)}(nS)$.

Figure~\ref{fig:pt-ratios} also displays the calculated full cross-section ratio between the $2^{++}$ and the $0^{++}$ ground states, $R = \sigma(pp\to T_{4c}(2^{++}(1S))+X)/\sigma(pp\to T_{4c}(0^{++}(1S))+X)$, with the masses, wave functions and $\alpha^2$ given in Table~\ref{tab:inputs}. When employing the values  from~\cite{Wang:2021kfv} ~(referred to as model A in Figure~\ref{fig:pt-ratios}) we obtain $R\approx 5\div 6$, which agrees well with a spin-counting rule. When employing the values from~\cite{Lu:2020cns}~(referred to as model B in Figure~\ref{fig:pt-ratios}) we obtain $R\approx 12\div 15$.

Table~\ref{tab:res} reports the integral values for the reduced cross-section $\overline{\sigma}$ at the collision energy $\sqrt{S} = 13~\text{TeV}$. The central values are calculated at the central scale $\mu = E_T$ with the average masses $\langle M(0^{++}) \rangle = \langle M(2^{++}) \rangle \simeq  6900~\text{MeV}$, $\langle M(1^{+-}) \rangle \simeq 6800~\text{MeV}$ and $\alpha^2 = 0.5$, which corresponds to an equal proportion of $[\bar 3\otimes 3]$ and $[6\otimes\bar 6]$ color configurations in the $0^{++}$ and  $0^{\prime ++}$ states. The ranges for the scale, mass, and $\alpha^2$ are the same as those used to estimate the uncertainties in the differential distributions. 
\begin{table}[ht]
\centering
\caption{Reduced cross-section $\overline{\sigma} = \sigma/\Phi$ at the proton-proton collision energy $\sqrt{S} = 13~\textrm{TeV}$ for particular kinematic conditions. For the definition and typical values of $\Phi$,~see Appendix~\ref{app:spectr_inputs}. The reduced cross-sections  are calculated with CTEQ18 parametrization for the strong coupling and gluon density functions. The values are presented along with their uncertainties as follows: $\overline{\sigma} + \Delta\overline{\sigma}(\text{scale}) + \Delta\overline{\sigma}(\text{mass})+ \Delta\overline{\sigma}(\alpha^2)$.}
\renewcommand{\arraystretch}{1.15}
\begin{tabular}{c||c|c|c}
\hline\hline
\multirow{3}{*}{Final state}
  & \multicolumn{3}{c}{Kinematic region} \\ \cline{2-4}
  & ~~~~$2 < y < 5$,\quad  $5 < p_T(\textrm{GeV}) < 50$~~~ & ~~~$|y| < 2$,\quad $10 < p_T(\textrm{GeV}) < 100$~~~ & ~~~$|y| < 2$,\quad $20 < p_T(\textrm{GeV}) < 100$~~~~   \\ 
 & (appr. corresponds to LHCb) & (accessible at CMS) & (accessible at ATLAS/CMS) \\ \hline
\multirow{2}{*}{$0^{++}(nS) + g$~} & \multirow{2}{*}{$550\ ^{+300}_{-200}\ ^{+40}_{-30}\ ^{+30}_{-20}$ pb} & \multirow{2}{*}{$110\ ^{+70}_{-40}\ ^{+7}_{-3}\ ^{+11}_{-4}$ pb} & \multirow{2}{*}{$3.70\ ^{+2.40}_{-1.40}\ ^{+0.30}_{-0.40}\ ^{+0.30}_{-0.30}$ pb}
\\ & & & \\ \hline
\multirow{2}{*}{$1^{+-}(nS) + g$~} & \multirow{2}{*}{$7.30\ ^{+2.60}_{-2.20}\ ^{+0.12}_{-0.02}$ pb} & \multirow{2}{*}{$0.90\ ^{+0.40}_{-0.30}\ ^{+0.07}_{-0.09}$ pb} & \multirow{2}{*}{$0.011\ ^{+0.006}_{-0.004}\ ^{+0.001}_{-0.002}$ pb}
\\ & & & \\ \hline
\multirow{2}{*}{$2^{++}(nS) +g$~} & \multirow{2}{*}{$7500\ ^{+4300}_{-2900}\ ^{+600}_{-550}$ pb} & \multirow{2}{*}{$1690\ ^{+950}_{-600}\ ^{+40}_{-70}$ pb} & \multirow{2}{*}{$57\ ^{+38}_{-21}\ ^{+5}_{-6}$ pb}
\\ & & & \\ \hline
\hline\hline
\end{tabular}
\label{tab:res}
\end{table}

Furthermore, we compared our cross-sections with those calculated in~\cite{Feng:2023agq}.
For the sake of a clear comparison, we repeated the calculation at the same  collision energy but in the kinematic region used in~\cite{Feng:2023agq}, $6 < p_T(\text{GeV})<100$, $|y| < 5$, and obtained the following values: $\overline{\sigma}(0^{++}) \approx 1.4~\text{nb}$,\; $\overline{\sigma}(1^{+-}) \approx 0.017~\text{nb}$, and $\overline{\sigma}(2^{++}) \approx 19.4~\text{nb}$. Our $\overline{\sigma}(1^{+-})$ and $\overline{\sigma}(2^{++})$ reduced cross-sections should be compared with the reduced cross-sections extracted from~\cite{Feng:2023agq} through  the conversion factor $10^{-10}\cdot 16\, M^9\,(2J+1)/\langle O^{(J)}_{3,3}\rangle$. In this way, we extract  $\overline{\sigma}(1^{+-}) \approx 0.17~\text{nb}$ and $\overline{\sigma}(2^{++}) \approx 104~\text{nb}$, which are larger than ours. In the case of the $0^{++}$ tetraquark, we are not able to extract  $\overline{\sigma}(0^{++})$ since the authors of~\cite{Feng:2023agq} matched the NRQCD matrix elements to the short-distance coefficients separately for each $[\bar 3\otimes 3]$ and $[6\otimes\bar 6]$ color configuration. Nevertheless, we notice that their full cross-section $\sigma(0^{++})$ is again larger than our estimation  $\Phi\,\overline{\sigma}(0^{++})\approx \Phi\, \cdot\, 1.4\, \text{nb}\,$, with 
$\Phi$  taken from Table~\ref{tab:inputs}.

It is worth assessing the impact of the IR regularization technique on our final results. The  $1^{+-}$ differential cross-section  $d\overline{\sigma}(1^{+-})/dp_T$ is finite at $p_T=0 $  and exhibits a physical shape which peaks at $p_T \simeq 1.5 $ GeV. By contrast, the $0^{++}$ and $2^{++}$ differential cross-sections obtained with a simple $p_T$ cut-off diverge at $p_T=0$. This means that, in the case of $0^{++}$ and $2^{++}$, we reproduce the physical $d\overline{\sigma}/dp_T$ shapes starting from some $p_T$, whereas 
we miss the peak at the lower $p_T$ expected in the physical shapes. All three differential cross-sections scale at large $p_T$ with a power law and exhibit a steeper rise in the middle $p_T$ range (approximately from 5 to 15 GeV). In particular, we found that, in the  $p_T> 15 $ GeV  region, the $d \overline{\sigma}/dp_T$ spectra can be fitted with the  simple power law functions:
\begin{align}
&   \frac{d\overline{\sigma}(J^{++})}{dp_T}  \sim \frac{a^{(J)}}{p_T^6(1+b^{(J)} p_T^2 )},  \quad J=\{0,2\},  
\label{eq:power-law-funcs_Jpp}
\\ 
&  \frac{d\overline{\sigma}(1^{+-})}{dp_T}  \sim \frac{a^{(1)}}{p_T^8(1+b^{(1)} p_T^2 )},
\label{eq:power-law-funcs_1pm}
\end{align}
where the coefficients for the central scale distributions shown in Figures~\ref{fig:pt-distrib-forward-b}, \ref{fig:pt-distrib-central-b} are given in Table~\ref{tab:fit_pars}.
\begin{table}[ht]
\centering
\caption{Coefficients for the $d \overline{\sigma}/dp_T$ spectra obtained by fitting the $p_T> 15 $ GeV region of the central scale distributions with the power law functions of Eqs.~\eqref{eq:power-law-funcs_Jpp},~\eqref{eq:power-law-funcs_1pm}. The fitting uncertainty is negligible.}
\renewcommand{\arraystretch}{1.25}
\begin{tabular}{|c||c|c|c|c|}
\hline \hline
\multirow{2}{*}{} & \multicolumn{2}{c|}{Acceptance $2 < y < 5$  }     & \multicolumn{2}{c|}{Acceptance $|y| < 2$}     
\\ \cline{2-5} 
& $a^{(J)}$  &  $b^{(J)}$ & $a^{(J)}$ &  $b^{(J)}$  
\\ \hline
~$J=0$~  & \hspace*{3ex} $3.816\cdot 10^7~\text{pb}\cdot\text{GeV}^5$ \hspace*{3ex} & \hspace*{3ex} $1.466\cdot 10^{-3}$ $\text{GeV}^{-2}$ \hspace*{3ex} & \hspace*{3ex} $7.602\cdot 10^7~\text{pb}\cdot\text{GeV}^5$ \hspace*{3ex}  &  \hspace*{3ex} $5.251\cdot 10^{-4}$ $\text{GeV}^{-2}$  \hspace*{3ex}     
\\ \hline
~$J=1$~  &  $4.169\cdot 10^7~\text{pb}\cdot\text{GeV}^7$ &     $1.039\cdot 10^{-3}$ $\text{GeV}^{-2}$  &   $1.140\cdot 10^{8}~\text{pb}\cdot\text{GeV}^7$    &   $3.035\cdot 10^{-4}$  $\text{GeV}^{-2}$       
\\ \hline
~$J=2$~  &  $5.773\cdot 10^8~\text{pb}\cdot\text{GeV}^5$   &   $1.382\cdot 10^{-3}$ $\text{GeV}^{-2}$ &  $1.175\cdot 10^9~\text{pb}\cdot\text{GeV}^5$     &  $5.081\cdot 10^{-4}$   $\text{GeV}^{-2}$      
\\ \hline \hline\hline
\end{tabular}
\label{tab:fit_pars}
\end{table}

The low $p_T$ cut-off, not smaller than 5 GeV, together with the applied acceptance restriction, ensures that we are far enough from the collinear divergent region, and thus can use our short-distance factors $F^{(0,2)}(s,t)$. In the middle $p_T$ range (approximately from 5 to 15 GeV) one might be concerned about overestimating the $0^{++}$ and $2^{++}$ cross-section magnitudes, owing to the proximity to the pole at $p_T=0$. However, we assume that the impact of this overestimation is subdominant compared to the large scale uncertainty typical of leading-order calculations. In conclusion, we suppose that the order of magnitude of the reduced cross-sections in Table~\ref{tab:res}  is trustworthy, and we are quite confident of their ratios.

\section{Comparison with the signal yields in the $J/\psi\, J/\psi$  spectrum}
\label{sec:yields}
The total yield of di-$J/\psi$ events measurable in the experiment includes a direct production component, a feed-down component (in which at least one $J/\psi$ originates from $\psi(2S)$, $\chi_c(1P)$ or other higher charmonium states), a resonant component, attributed to an exotic fully charmed state, and a small rate of non-prompt events from $b$-hadron decays. It should be observed that the prompt nonresonant di-$J/\psi$ yields~(direct and feed-down) are given by the sum of the contributions of the SPS and DPS production mechanisms. By contrast, the resonant production is exclusively associated with the SPS mechanism.

Assuming that the resonantly produced $J/\psi$-pairs originate from a $T_{4c}$ tetraquark, we can pin down a few possible resonant contributions, which we refer to as the direct, the charmonium feed-down and the tetraquark feed-down contributions. The direct component corresponds to a $T_{4c}$ directly produced and directly decaying into the $J/\psi$-pair. It can be observed in the fully reconstructed decay, and its yield can be estimated as $N_\textrm{direct}^\textrm{(SPS)}(T_{4c})\cdot\textrm{Br}(J/\psi\,J/\psi)$. The charmonium feed-down component corresponds to a $T_{4c}$ directly produced and decaying into the $J/\psi$-pair via at least one intermediate higher charmonium state. As an example of the charmonium feed-down component, one might consider the $T_{4c}\to \chi_c(\to J/\psi\,\gamma)\:\chi_c(\to J/\psi\,\gamma)$ partially reconstructed decay chain, where the photons escape the reconstruction. The yield of such a contribution, whose decay occurs via the $T_{4c}\to {\cal Q}(\to J/\psi\,X)\:{\cal Q}(\to J/\psi\,X)$ channel~(${\cal Q}$ is the higher charmonium state, and $X$ is the missing decay product) can be estimated as  $N_\textrm{direct}^\textrm{(SPS)}(T_{4c})\cdot\textrm{Br}({\cal Q}\,{\cal Q})\cdot\textrm{Br}^2({\cal Q}\to J/\psi\,X)$. The tetraquark feed-down component corresponds to a $T_{4c}$ that proceeds from a higher mass tetraquark decay and subsequently decays into the $J/\psi$-pair. As an example of the tetraquark feed-down component one might consider the  $T_{4c}(2S)\to T_{4c}(1S)(\to J/\psi\,J/\psi)\,\pi^+\pi^-$ decay chain.

The total yield of di-$J/\psi$ events measurable in the experiments can be written down as
\begin{multline}\label{yield_total}
    N_\textrm{tot}(J/\psi\,J/\psi) = N_\textrm{direct}^\textrm{(SPS+DPS)}(J/\psi\,J/\psi) + N_\textrm{feed-down}^\textrm{(SPS+DPS)}(J/\psi\,J/\psi) + \left(N_\textrm{direct}^\textrm{(SPS)}(T_{4c}) + N_\textrm{feed-down}^\textrm{(SPS)}(T_{4c})\right)\cdot\textrm{Br}(J/\psi\,J/\psi) \\+ N_\textrm{direct}^\textrm{(SPS)}(T_{4c})\cdot\textrm{Br}({\cal Q}\,{\cal Q})\cdot\textrm{Br}^2({\cal Q}\to J/\psi\,X) + N_\textrm{non-prompt}(J/\psi\,J/\psi)  + \ldots
    \,,
\end{multline}
where the  resonant contribution is proportional to $ N_\textrm{direct}^\textrm{(SPS)}(T_{4c})$ and $N_\textrm{feed-down}^\textrm{(SPS)}(T_{4c})$, and can be expanded with many other decay chains occurring in the charmonium and mixed charmonium-tetraquark feed-down.

In this subsection, we estimate only the yields of the main resonant component, $N_\textrm{direct}^\textrm{(SPS)}(T_{4c})\cdot\textrm{Br}(J/\psi\,J/\psi)$, in order to draw some conclusions from its comparison with the measured yields. In that respect, let us review the available data on the signal yields in the di-$J/\psi$ spectrum. First, in all three researches~\cite{LHCb:2020bwg,ATLAS:2023bft,CMS:2023owd} the narrow peak around 6900 MeV has been observed. Alongside the clear peak near  6900 MeV a relatively broad structure, starting from $6200~\text{MeV}$~(the di-$J/\psi$ threshold), and a narrow structure at a higher invariant mass around 7300 MeV have been seen~\cite{LHCb:2020bwg,ATLAS:2023bft,CMS:2023owd}. While the shape of the first broad structure in the di-$J/\psi$ mass spectrum appears consistent across all three studies, and it certainly indicates the presence of the experimental excess over the background, the yield of the third structure is suppressed and differs across the three experiments. In the di-$J/\psi$ mass spectrum measured by LHCb~\cite{LHCb:2020bwg}, the small bump seen just below 7300 MeV was neglected because of the low statistical significance. No evidence of this structure was found later by ATLAS~\cite{ATLAS:2023bft}. In the recently published CMS study~\cite{CMS:2023owd} this structure was confirmed at 7287 MeV with a statistical significance of $4.7\,\sigma$.

Before making comparisons with the measured yields, it is important to point out some details in fitting the mass spectrum in~\cite{LHCb:2020bwg,ATLAS:2023bft,CMS:2023owd}. So far, the experimental publications assume a fixed shape for the background, based on pQCD, and coherent or incoherent sum of Breit-Wigners to describe the resonant peaks. Both the non-interfering and interfering with background scenarios were considered in fitting the mass spectrum.

Concerning the first broad structure, the statistical significance in all three experiments exceeds $5.0\,\sigma$. In the CMS study~\cite{CMS:2023owd} this broad structure was resolved into a single peak and an additional component, this latter being responsible for the threshold enhancement just above 6200 MeV due to the opening of the di-$J/\psi$ threshold.  This additional component was incorporated into the background fit by introducing a separate  Breight-Wigner function. By contrast with CMS, in the LHCb study~\cite{LHCb:2020bwg} the full broad structure in the mass region $6200\div 6700~\text{MeV}$ was treated as a threshold enhancement. The yield enhancement in this region was fitted with two Breight-Wigner functions, called ``Threshold BW$_1$'' and ``Threshold BW$_2$'' components in~\cite{LHCb:2020bwg}. Although these two Breight-Wigners could fit the overlap between the threshold enhancement and a possible genuine resonant state, the broad structure was not interpreted as incorporating a true resonance. For this reason, no yield of the first broad structure was published by LHCb in~\cite{LHCb:2020bwg}. In the ATLAS study~\cite{ATLAS:2023bft} the first broad structure was fitted with two interfering Breight-Wigner functions. This interference hinted at the possibility that the di-$J/\psi$ mass spectrum in the region $6200\div 6700~\text{MeV}$ might have a more complex structure, comprising two coherently produced states with the same quantum numbers. It is important to add that, in ATLAS, the feed-down events --- the di-$J/\psi$ events with one of the $J/\psi$ coming from the $\psi(2S)$ cascade decays --- were included as an additional background. This charmonium feed-down contribution alters the background shape in the region of the first broad structure and, particularly, results in a small bump just above the di-$J/\psi$ threshold; this could be one of the explanations for the threshold enhancement seen in all three experiments.

A few remarks on interpreting the interference should be made. On investigating the di-$J/\psi$ mass spectrum, LHCb and ATLAS also applied fitting models that allowed for the interference between the SPS component of the background and a Breight-Wigner for the threshold enhancement. While in the experimental analysis such a model is able to improve the quality of fit, it is difficult to say if such fit supports the hypothesis that the first broad structure includes a genuine resonance.\,\footnote{\label{interference}
This type of interference was considered to achieve a better agreement with data in the region of the dip between the first broad structure and the peak around $6900$ MeV. We suggest that the SPS component of the background (as well as the DPS one) represents an incoherent production of di-$J/\psi$ pairs in different configurations with various $J^{PC}$. However, any background might also contain a fraction of $S$-wave $J/\psi$-pairs that will interfere with the Breit-Wigners representing genuine $0^{++}$ or $2^{++}$ resonant state. What is this fraction and how it depends on the invariant mass, it can be established only from the experimental angular distributions.} On interpreting these fitting models, we tend to consider the $T_{4c}$ resonance only the peak around 6900 MeV and relate the entire first broad structure to the background. In the CMS study~\cite{CMS:2023owd} the models taking into account the interference with the background did not improve the fit quality. Thus, CMS applied a different fitting model, one in which all three observed resonances interfere with each other. Incorporating these interference terms showed a significant improvement in the regions of the dips between the resonances. We note that this fitting model unavoidably requires all three peaks to be assigned to the same  $J^{PC}$ quantum numbers.

The amount of data on the measured yields is rather poor. ATLAS did not report any signal yields~\cite{ATLAS:2023bft}, and CMS reported  only the signal yields measured within the ``no interference'' fitting model~\cite{CMS:2023owd}:
\begin{equation}\label{CMS-yields}
\begin{array}{ccccc}
N_\textrm{CMS}(6552) = 470^{+120}_{-110},\qquad\qquad
&N_\textrm{CMS}(6927) = 492^{+78}_{-73},\qquad\qquad
&N_\textrm{CMS}(7287) = 156^{+64}_{-51},
\end{array}
\end{equation}
where the central value of the measured mass is written in parentheses.  LHCb reported  only the signal yield of $X(6900)$ in the ``$p_T^{\text{di-}J/\psi}$-threshold'' approach~\cite{LHCb:2020bwg}. Two values can be exploited: one obtained from the ``no interference'' model, and one obtained from the model with interference between the Breight-Wigner for the threshold enhancement and the SPS background component
(denoted with a tilde): 
\begin{equation}\label{LHCb-yields}
\begin{array}{ccccc}
&N_\textrm{LHCb}(6905) = 252\pm 63,\qquad\qquad& 
&\widetilde{N}_\textrm{LHCb}(6886) = 784\pm 148,\qquad\qquad& 
\end{array}
\end{equation}
where the central value of the measured mass is written in parentheses. 
Of the three experiments, only LHCb provided the experimental efficiency $\varepsilon_\textrm{LHCb} = 0.09$~(see supplementary material D in the arXiv version~\cite{LHCb:2020bwg}).

It should be remembered that the quoted yields for each di-$J/\psi$ structure heavily depend on the fitting procedure.  First of all, there is no justification that pQCD predicts the correct lineshape of the background so close to hadron threshold, where final state interaction can play a major role. Furthermore, to assess whether the Breit-Wigner sum is coherent or incoherent, one would need to extract first the quantum numbers from the angular distributions. In lack of all this information, the yields~\eqref{CMS-yields},\eqref{LHCb-yields} are subject to large systematic uncertainties.

The experimental yields~\eqref{LHCb-yields} can be quantitatively compared with the estimated signal yields of the main resonant component 
in Eq.~\eqref{yield_total},
\begin{equation}
    \label{sig-yield}
N_{\textrm{sig}}\approx\varepsilon{\cal L}\,\Phi^{}\,\overline{\sigma}(pp\to~T_{4c}+X)\cdot \text{Br}(T_{4c}\to J/\psi\,J/\psi),
\end{equation}
for both $T_{4c}(0^{++})$ and $T_{4c}(2^{++})$. Here, ${\cal L}  $  denotes the integral luminosity at Run~I and Run~II collision energies, so that  ${\cal L}\overline{\sigma}~=~\overline{\sigma}(8~\text{TeV})\cdot 3~\text{fb}^{-1} + \overline{\sigma}(13~\text{TeV})\cdot 6~\text{fb}^{-1}$.
The reduced cross-sections for LHCb  at 
13 TeV are given in  Table~\ref{tab:res}, while their central values at  8 TeV are
$\overline{\sigma}(0^{++}) = 310~\text{pb}$ and  $\overline{\sigma}(2^{++}) =4200~\text{pb}$.
As mentioned in~\cite{LHCb:2020bwg}, the experimental efficiency $\varepsilon_\textrm{LHCb}$ was found to be marginal across the investigated range of di-$J/\psi$ mass. However, no comments were made on the efficiency stability across the range in~$p_T^{\text{di-}J/\psi}$. We presume that it is sufficiently stable at least at $p_T^{\text{di-}J/\psi} = (5\div 10)~\text{GeV}$,
where
our calculated cross-section reaches 90\,\% of its integral value.

A direct  estimation of $N_{\textrm{sig}}$ using $\text{Br}(T_{4c}\to J/\psi\,J/\psi) \sim 10^{-3}$ as estimated in~\cite{Becchi:2020uvq}, and the wave functions from Appendix~\ref{app:spectr_inputs} gives
\begin{align}
& N_{\textrm{sig}}(0^{++}) \sim 100 \div 1000\, ,
& & N_{\textrm{sig}}(2^{++}) \sim 1000 \div 10000\,,  & \text{for $1S$}, &  \nonumber \\
\label{yields_estimations}
& N_{\textrm{sig}}(0^{++}) \sim 10 \, ,
& & N_{\textrm{sig}}(2^{++}) \sim 100 \, ,  & \text{for $2S$}, &
  \\ 
& N_{\textrm{sig}}(0^{++}) \sim 10,
& & N_{\textrm{sig}}(2^{++}) \sim 100 \, , &  \text{for $3S$}, & \nonumber
\end{align}
where the large indetermination for the $1S$ states is because the $\Phi\equiv \Phi^{(J)}(nS)$ reduced wave functions calculated in~\cite{Wang:2021kfv} and~\cite{Lu:2020cns} differ by  up to one order of magnitude. We point out that the $\text{Br}(T_{4c}\to J/\psi\,J/\psi)$ branching ratios are known with even higher ambiguity. For example, one can find predicted values ranging from $10^{-3}$~\cite{Becchi:2020uvq} to 1~\cite{Sang:2023ncm}.

Because of this large indetermination, it is reasonable to use the experimental yields to constraint the product  $\Phi^{}(nS)\cdot\text{Br}(T_{4c}\to J/\psi\,J/\psi)$ in the scenario in which $X(6900)$ is either a $0^{++}$ or $2^{++}$ fully charmed tetraquark. We equalize the estimation~\eqref{sig-yield} to the untilded yield in Eq.~\eqref{LHCb-yields} and derive
\begin{align}\label{K_times_Br}
& \Phi^{(0)}(nS) \cdot\textrm{Br}(T_{4c}(0^{++})) = (0.5\div 1.7)\cdot 0.662\cdot 10^{-3},
& \Phi^{(2)}(nS) \cdot\textrm{Br}(T_{4c}(2^{++})) = (0.5\div 1.7)\cdot 0.049\cdot 10^{-3},
\end{align}
where the factor $(0.5\div 1.7)$ in front reflects the sum of the theoretical (coming mainly from $\overline{\sigma}$ dependence on scale) and  experimental uncertainties. If one takes the tilded value from~\eqref{LHCb-yields}, the right-hand side of the derived relations will be approximately three times larger. However, in our opinion,this means 
rejecting the possibility of a genuine resonance in the first broad structure in the di-$J/\psi$ mass spectrum.

We observe that Eq.~\eqref{K_times_Br} is a new model-independent constraining criterion made possible by our pQCD calculations. Each derived range reflects uncertainties in our calculations and in the yield measured under a specific fitting model. However, one should keep in mind that the obtained ranges may differ significantly if a signal yield from a different fitting procedure is used in the derivation.

\section{Discussion}
\label{sec:disc}
\subsection{The $X(6900)$}
\begin{figure}[ht]
  \centering
  \resizebox*{0.4\textwidth}{!}{\includegraphics{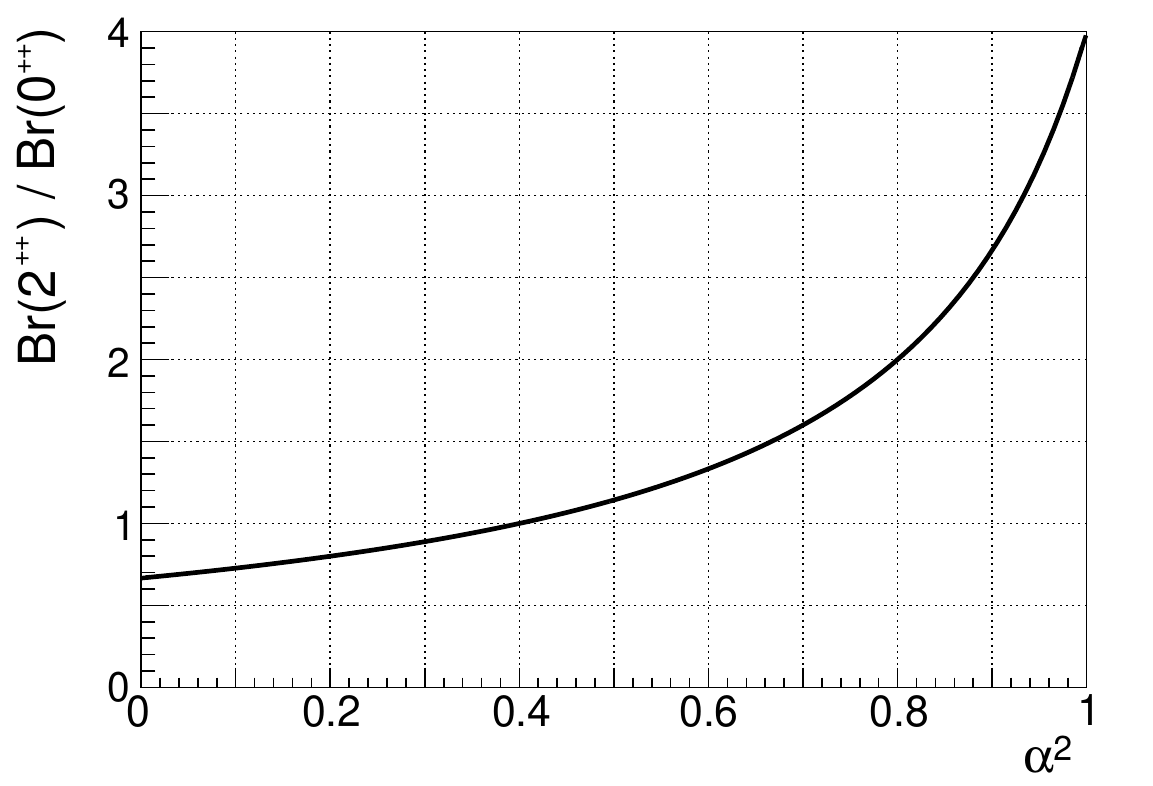}}
\caption{Naive estimation of the $\text{Br}\left(T_{4c}(2^{++})\to J/\psi\,J/\psi)\right)/\text{Br}\left(T_{4c}(0^{++})\to J/\psi\,J/\psi)\right) $  ratio as a function of the weight of $[\bar 3\otimes 3]$ color configuration in $T_{4c}(0^{++})$. The endpoints are $\alpha^2=1$~(pure $[\bar 3\otimes 3]$ color state) and $\alpha^2=0$~(pure $[6\otimes\bar 6]$ color state).}
\label{fig:Brs_ratio}
\end{figure}

It is straightforward to check whether the predicted $T_{4c}$ reduced wave functions and branching ratios fulfill the constraints~\eqref{K_times_Br}. In contrast to the relative confidence in the production mechanism given by the leading-order pQCD calculation in CSM, the values of branching ratios are model-dependent. The branching ratios can be extracted from~\cite{Becchi:2020uvq}, where the authors applied an effective decay model based on the Vector Meson Dominance~(VMD) and symmetry principles. The $\textrm{Br}(T_{4c}(0^{++})\to J/\psi\,J/\psi)$ branching ratio was obtained from the VMD estimation, and the $\textrm{Br}(T_{4c}(2^{++})\to J/\psi\,J/\psi)$ branching ratio was obtained from the former multiplying by the simple scaling factor $\frac{1/3}{\left[\alpha^2/12+ (1-\alpha^2)/2\right]}$, which is the ratio of relative weights of the $\big||c\bar c \rangle_1^{\;1}\otimes |c\bar c \rangle_1^{\;1}\,\rangle_1^{\;0}$  components in the meson-like basis~(see the coefficients in~\eqref{main_spin0} and~\eqref{main_spin2}). The authors of~\cite{Becchi:2020uvq} considered only the $[\bar 3\otimes 3]$ color configuration in $T_{4c}(0^{++})$. Consequently, the scaling factor was taken at $\alpha^2=1$ and amounted to 4,
\begin{align}\label{branchings}
& \textrm{Br}(T_{4c}(0^{++})\to J/\psi\,J/\psi) \approx 0.73\cdot 10^{-3}, 
& \textrm{Br}(2^{++})/\textrm{Br}(0^{++})= 4.
\end{align}
In the present study, we consider the generic case in which $T_{4c}(0^{++})$ is a superposition of both the $[\bar 3\otimes 3]$ and the $[6\otimes\bar 6]$ color configurations;  consequently the branching ratio  of the $2^{++}$ state should be adjusted accordingly
\begin{align}\label{branchings_generic}
& \textrm{Br}(T_{4c}(0^{++})\to J/\psi\,J/\psi) \approx 0.73\cdot 10^{-3}, 
& \textrm{Br}(2^{++})/\textrm{Br}(0^{++})= \frac{1/3}{\left[\alpha^2/12+ (1-\alpha^2)/2\right]}. 
\end{align}
The plot of $\textrm{Br}(2^{++})/\textrm{Br}(0^{++})$ as a function of the weight $\alpha^2$~($0\leq \alpha^2 \leq 1$) is shown in Figure~\ref{fig:Brs_ratio}. One can clearly see that the factor 4 remains an upper value for $\textrm{Br}(2^{++})/\textrm{Br}(0^{++})$, and the presence of the $[6\otimes\bar 6]$ color state reduces this ratio down to its minimal value 2/3~(only $[6\otimes\bar 6]$ color state).

Table~\ref{tab:predictions} shows the product of the branching  ratios~\eqref{branchings_generic} and the relevant reduced wave functions (listed in  Appendix~\ref{app:spectr_inputs}) for the potential candidates for $X(6900)$. Since in both spectroscopy models investigated in~\cite{Wang:2021kfv,Lu:2020cns}, the $[6\otimes\bar 6]$ color configuration is present, the spin-zero state is degenerated to two states with close masses, $0^{++}$ and $0^{\prime\,++}$. 
Therefore, in Table~\ref{tab:predictions} we treat $0^{++}(2S)$ and $0^{\prime\,++}(2S)$  as two distinct states. In the measurements, these states can potentially overlap with each other giving a peak in the mass range of the $(0^{++},\,0^{\prime++})$ doublet. Considering this scenario one should simply account for the sum of $\Phi^{(0)}(2S)\cdot\textrm{Br}(T_{4c}(0^{++}))$ and $\Phi^{(0^\prime)}(2S)\cdot\textrm{Br}(T_{4c}(0^{++}))$.

\begin{table}[ht]
 \centering
\caption{Predicted parameters of $T_{4c}$ candidates for $X(6900)$. The values should be compared with Eq.~\eqref{K_times_Br}.}
\label{tab:predictions}
\renewcommand{\arraystretch}{1.45}
    \begin{tabular}{|c|c|c|c|c|c|c|}
    \hline
    \hline
    & \multicolumn{2}{|c|}{$0^{\,++}(2S)$} & \multicolumn{2}{|c|}{$0^{\prime ++}(2S)$}   & \multicolumn{2}{|c|}{$2^{++}(2S)$} \\ \cline{2-7}
   & Mass, MeV & $\Phi^{(0)}(2S)\cdot\textrm{Br}(T_{4c}(0^{++}))$ & Mass, MeV & $\Phi^{(0^\prime)}(2S)\cdot\textrm{Br}(T_{4c}(0^{++}))$ & Mass, MeV & $\Phi^{(2)}(2S)\cdot\textrm{Br}(T_{4c}(2^{++}))$  \\ \hline
Refs.~\cite{Becchi:2020uvq,Wang:2021kfv}~ & $6867 $ & $0.047\cdot 10^{-3}$ & 7007 & $0.091\cdot 10^{-3}$ &  6917 & $0.034\cdot 10^{-3}$
\\ \hline
Refs.~\cite{Becchi:2020uvq,Lu:2020cns}~  & $6849 $ & $0.059 \cdot 10^{-3}$ & 6940 & $0.070 \cdot 10^{-3}$ & 6948 & $0.109\cdot 10^{-3}$
\\ \hline
    \hline
    \hline
\end{tabular}
\end{table}

Table~\ref{tab:predictions} reveals that the comparison with the  constraints~\eqref{K_times_Br} is in favour of the $2^{++}(2S)$ hypothesis. The $\Phi\cdot\textrm{Br}$ values predicted for the $0^{++}(2S)$ and $0^{\prime++}(2S)$ states are a few times smaller than the lower limit of~\eqref{K_times_Br}, whereas the $\Phi\cdot\textrm{Br}$ values predicted for the $2^{++}(2S)$ state either fulfill the derived constraint or slightly exceed its upper limit. This agrees with the fact that the $0^{(\prime) ++}(2S)$  yields estimated in the LHCb conditions are found to be notably smaller than $N_\textrm{LHCb} \approx 252$, as shown in Eq.~\eqref{yields_estimations}. 
Of course, the values from Table~\ref{tab:predictions} are not drastically different from each other and might be all compatible with the constraints if one allows for the large uncertainties of the fitting model itself. Nonetheless, we do our best to discriminate between the spin-zero and spin-two hypotheses
making use of the available measurements. In summary, taking into account the spin-two enhancement both in production~(see Figure~\ref{fig:pt-ratios}) and  decay~(see Figure~\ref{fig:Brs_ratio}), and the fulfilment of the constraining criteria~\eqref{K_times_Br}, we find $2^{++}(2S)$ to be the best matching $T_{4c}$ candidate for $X(6900)$.

It is important to stress the consistency in the orders of magnitude among the three independently predicted quantities: the pQCD cross-sections calculated in this article, the four-body wave functions from the nonrelativistic quark models~\cite{Wang:2021kfv,Lu:2020cns} and the branching ratios estimated with the help of VMD model~\cite{Becchi:2020uvq}. These low branching ratios contradict several phenomenological studies on $T_{4c}$ decays, in which the hidden flavor decay channels are assumed to saturate the total $T_{4c}$ width~\cite{Sang:2023ncm,Wang:2023kir,Agaev:2023wua}. Indeed, all those predictions assume that the other decay channels, such as $D^{(*)}\,\bar{D}^{(*)}$ or $J/\psi(\eta_c) + X$, are not significant, and that the di-$J/\psi$ and di-$\eta_c$ modes together saturate the $C$-even $T_{4c}$ decay widths (note also the $\Gamma_\text{tot}\approx \Gamma(J/\psi\,J/\psi)$ saturation in~\cite{Sang:2023ncm}). As a result, the branching ratios calculated in~\cite{Sang:2023ncm,Wang:2023kir,Agaev:2023wua},
 $\text{Br}(T_{4c}\to J/\psi\,J/\psi)\approx (0.5\div 1)$, lead to a three-order-of-magnitude discrepancy. By contrast, the good order of magnitude agreement for the values of Eq.~\eqref{branchings_generic} hints that di-$J/\psi$ decay mode is suppressed~\cite{Becchi:2020uvq}.

\subsection{Interpretation of the remaining two di-$J/\psi$ structures}
In addition to the outlined interpretation of the $X(6900)$ state, the remaining two structures in the di-$J/\psi$ spectrum should be addressed. The yields of the first two peaks observed by CMS are approximately in a $1:1$ ratio. As discussed earlier, we expect the $2^{++}$ dominance in the $T_{4c}$ visibility factors because the combined $2^{++}$ enhancement factor in production and decay is much larger than the corresponding suppression factor in the wave function. This finding disproves the hypothesis of the $N(0^{++}):N(2^{++})\approx 1:1$ ratio for the first two peaks observed by CMS. The spectroscopy models~\cite{Wang:2021kfv,Lu:2020cns} do not predict the ratio $\Phi^{(2)}(1S):\Phi^{(2)}(2S)$ to be close to $1:1$, which could otherwise explain the first two peaks as the $2^{++}(1S)$ and $2^{++}(2S)$ states. In our opinion, the hypothesis of the $T_{4c}(2^{++}(1S)):T_{4c}(2^{++}(2S))$ assignment deserves to be considered, but more precise studies of the four-body wave functions and the di-$J/\psi$ branching dependence on the radial state are needed in order to work out the constraints on the  $\Phi^{(2)}(1S)\:\text{Br}(1S):\Phi^{(2)}(2S)\:\text{Br}(2S)$ ratio.

Most likely, in LHCb and ATLAS, the broad structure in the mass region $6200\div 6700~\text{MeV}$ represents an overlap of the threshold enhancement contribution and one or more genuine resonance(s). Given the latest CMS results, where the interference with the background is unfavorable to the fitting model, the scenario in which the entire broad structure manifests as the threshold effect seems less significant. Besides the directly produced $T_{4c}(2^{++}(1S))$ the di-$J/\psi$ broad structure might include several suppressed contributions. The subleading contributions are expected to come from the directly produced $0^{++}(1S)$ tetraquark and from the $2^{++}(1S)$ tetraquark produced through the  feed-down from $2^{++}(2S)$ in the $T_{4c}(2^{++})\to T_{4c}(2^{++})\left\lbrace\pi^0\pi^0,\pi^+\pi^-\right\rbrace$ hadronic mode. In both cases, we would assume a one-order-of-magnitude suppression in the visibility factor. In addition to the fully reconstructed $T_{4c}$ states, the di-$J/\psi$ broad structure contains contributions from the partially reconstructed $T_{4c}$ decay products that may impact the shape of the mass spectrum. For example, consider the $T_{4c}(2^{++})\to J/\psi\:\psi(2S)(\to J/\psi \left\lbrace\pi^0\pi^0,\pi^+\pi^-\right\rbrace)$ or $T_{4c}(2^{++})\to \chi_c(\to\gamma\, J/\psi)\:\chi_c(\to\gamma\, J/\psi)$ decay modes.

The enhancement in the di-$J/\psi $ mass spectrum around 7300~MeV exceeds the predicted masses of the $3S$ radial excitations given in Table~\ref{tab:inputs} by 250~MeV, making it challenging to attribute this peak~(bump) to an $S$-wave tetraquark.   
In Ref.~\cite{Karliner:2020dta} Karliner and Rosner observed that the di-$J/\psi $ invariant mass region between 6200 and 7300~MeV is rich in meson-meson and baryon-baryon thresholds that might be responsible for such a complex signal structure. As was observed, the dip around $6750\,\mathrm{MeV}$ might correspond to the opening of the $S$-wave di-$\chi_{c0}$ channel at $\simeq 6829$~MeV, while the dip around $7200\,\mathrm{MeV}$ could be correlated with the opening of the di-$\eta_c(2S)$ channel at $\simeq 7276$~MeV or $\Xi_{cc} \bar{\Xi}_{cc}$ channel at $\simeq 7244$~MeV. Thus, the observation of the bump at 7300 MeV might only be induced by the opening of the di-$\eta_c(2S)$ or   $\Xi_{cc} \bar{\Xi}_{cc}$ channels. Combined information of both the four-body calculation mass predictions~\cite{Wang:2021kfv,Lu:2020cns} and exploration of the relevant thresholds in the di-$J/\psi $ invariant mass~\cite{Karliner:2020dta}, prompts us to avoid interpreting the third peak (bump) as a genuine tetraquark state.

\section{conclusions}
\label{sec:concl}
This article presents the factorization formalism for the production of an $S$-wave fully heavy tetraquark in the SPS interaction. The formalism is based on the Fierz transformation, which allows one to construct the production amplitudes in the meson-like basis, for which the spin-color projection techniques are well established. The  formalism outlined is applied to the calculation of $T_{4c}(0^{++},1^{+-},2^{++})$ hadronic production cross-sections. The cross-sections are predicted in the kinematic regions accessible at the LHC experiments.

In the collinear approach, the leading-order process of  $T_{4c}$ production with experimentally observable transverse momentum necessarily implies the $2\to 2$ scattering process. Therefore, we calculate the inclusive $\sigma(pp\to T_{4c}+X)$ cross-sections in the dominant channel of gluon fusion at the leading order, ${\cal O}(\alpha_s^5)$. The ${\cal O}(\alpha_s^5)$ short-distance factors are obtained from the pQCD calculations of the SPS matrix elements, projected onto all allowed $T_{4c}$ internal color and spin configurations compatible with the symmetry principles. The short-distance factors are further combined with the four-body $T_{4c}$ wave functions at the origin. The partonic cross-sections obtained are convoluted with the CTEQ18 gluon density functions and integrated over the $(p_T,\eta)$ ranges accessible in the ATLAS, CMS and LHCb detectors.

We observe a few orders of magnitude suppression in the $T_{4c}(1^{+-})$ production rate as well as dominance of the $T_{4c}(2^{++})$ production rate relative to that of $T_{4c}(0^{++})$. We predict the integral values and $p_T$-distributions of the reduced cross-sections $\overline{\sigma} = \sigma/\Phi$, with the normalization given by the dimensionless factor $\Phi = 10^{10}\,|\Psi(0,0,0)|^2/M^9$ derived from the four-body spectroscopy models~(see Appendix~\ref{app:spectr_inputs}). In the large $p_T$ limit, the differential distributions scale as $d\overline{\sigma}(0^{++},2^{++})/d p_T \sim 1/p_T^6$ ~and~ $d\overline{\sigma}(1^{+-})/d p_T \sim 1/p_T^8$. The ratio  $r = \overline{\sigma}(2^{++})/\overline{\sigma}(0^{++})$ is found to be insensitive to the acceptance range, slightly increasing with $p_T$ and reaching saturation in the large $p_T$ limit. For the $5 \le p_T(\textrm{GeV})\le 100$ cut, the ratio lies in the $12\le r(p_T)\le 17$ range.

Despite some differences in the fitting models, there is a clear consensus among ATLAS, CMS and LHCb regarding the existence of $X(6900)$ in the di-$J/\psi$ invariant mass spectrum. Thanks to the fact that LHCb has published both the fiducial cuts on $p_T^{\textrm{di-}J/\psi}$ and the experimental efficiency, we could compare the estimated number of $pp \to T_{4c}(\to J/\psi\,J/\psi) +X$ events with the $X(6900)$ signal yield at LHCb. Most important, we derive the following model-independent constraints for the  $T_{4c}$ candidates for $X(6900)$:
\begin{itemize}
\item $0^{++}(nS)$ states \\
$0.331\cdot 10^{-3} \leq \,\Phi^{(0)}(nS) \cdot\textrm{Br}(T_{4c}(0^{++})\to J/\psi\,J/\psi) \,\leq   1.125\cdot 10^{-3}$ ;
\item $2^{++}(nS)$ states \\
$ 0.025\cdot 10^{-3} \leq \,  \Phi^{(2)}(nS) \cdot\textrm{Br}(T_{4c}(2^{++})\to J/\psi\,J/\psi)\,\leq  0.083 \cdot 10^{-3}$ .
\end{itemize}

These constraints are checked for the $T_{4c}$ states by using  the four-body mass spectrum calculations~\cite{Wang:2021kfv,Lu:2020cns} and the branching ratios from~\cite{Becchi:2020uvq} updated to include the $\left[6 \otimes \bar 6\right]$  color configuration for the $T_{4c}(0^{++})$ state. We observe an order of magnitude consistency for $\Phi^{(J)}(nS)\sim (0.01\div 0.1)$  and  $\textrm{Br}(T_{4c}\to J/\psi\,J/\psi)\sim 10^{-3}$. Considering the two predicted $T_{4c}$ states with masses close to 6900~MeV, namely $0^{++}(2S)$ and $2^{++}(2S)$, we find the $2^{++}(2S)$ state to be the best matching candidate for $X(6900)$.

In the course of future searches extending the first investigations~\cite{LHCb:2020bwg,ATLAS:2023bft,CMS:2023owd}, it will be useful to measure the cross-sections of the resonant di-$J/\psi$ component (scaled by the unknown branching ratio), such as, for example, $\sigma(pp\to X(6900))\cdot\text{Br}(J/\psi\,J/\psi)$. The comparison of the measured cross-sections with the theoretical predictions can probe the tetraquark nature of the structures observed  in the di-$J/\psi$ mass spectrum. A good agreement with the cross-sections obtained by multiplying our results of Table~\ref{tab:res} and the model-dependent values of $\Phi^{(J)}(nS)$  and  $\text{Br}(J/\psi\,J/\psi)$ will hint at a tetraquark nature. In this regard, we encourage experimentalists in ATLAS, CMS and LHCb to conduct the cross-section measurement for the resonant di-$J/\psi$ component in the fiducial $(p_T^{\textrm{di-}J/\psi},\: \eta^{\textrm{di-}J/\psi})$ volume.

As a final comment, we note that the NLO corrections to the cross-sections might give a sizeable contribution and help mitigate the large scale uncertainty. 
We also remark that the $T_{4c}$ production mechanism can be studied via other approaches, for example, by means of $k_T$-factorization exploiting the short-distance factors of the order of ${\cal O}(\alpha_s^4)$. Finally, the impact of color-octet contributions and production in the DPS interaction can be investigated.

\acknowledgments
We are pleased to express our gratitude to Profs. C.~M.~Becchi and L.~Maiani for their encouragement, valuable discussions and comments on the first version of the manuscript. We also would like to thank Prof. A.~van~Hameren for informative discussions on calculation methods. A.~G. acknowledges support from the CIDEGENT program with Ref.~CIDEGENT/2019/015, the Spanish Ministerio de Economia y Competitividad and European Union (NextGenerationEU/PRTR) by the grant with Ref. CNS2022-1361.

\newpage
\appendix
\section{Spectroscopy inputs}
\label{app:spectr_inputs}
Refs.~\cite{Wang:2021kfv}~and~\cite{Lu:2020cns} publish results on the $T_{4c}$ mass spectrum in the nonrelativistic four-body quark model. However, we observe that in these studies the numerical values of the wave functions are not reported. Nevertheless, the wave functions calculated at the origin can be extracted from the root mean squared radii in a conventional way. Table~\ref{tab:inputs} collects together the percentages of the two color configurations,  masses, and root mean square radii for the $T_{4c}$ mass spectra from~\cite{Wang:2021kfv,Lu:2020cns}, along with the extracted wave functions at the origin.

\begin{table}[ht]
 \centering
 \vspace{1ex}
\caption{From left to right: percentage of color configurations, mass in GeV,  modulus squared of the four-body wave function at the origin in GeV$^9$,  reduced wave function $ \phi = |\Psi(0,0,0)|^2/M^9$, and root mean squared radii in fm.}
   \label{tab:inputs}
   \begin{ruledtabular}
    \begin{tabular}{c|c|ccccccccc}
\multicolumn{2}{c|}{} & $|\bar 3\otimes 3\rangle$ & $~|6\otimes\bar 6\rangle$ & Mass &  $10^3|\Psi(0,0,0)|^2$ & $10^{10} \phi$ & $\sqrt{\langle r^2_{12(34)} \rangle}  $&  $\sqrt{\langle r^2 \rangle}  $ \\ 
\hline
\multirow{10}{*}{Ref.~\cite{Wang:2021kfv}~} & $0^{\,++}(1S)$  & 31.9\% & 68.1\% & 6.405  & 1.20 & 0.661 & 0.52 & 0.31  \\
&$0^{\prime++}(1S)$  & 67.7\% & 32.3\% & 6.498  & 0.88 & 0.426 & 
0.51 & 0.36  \\
&$0^{\,++}(2S)$  & 10.6\% & 89.4\% & 6.867   &  0.22 & 0.065 & 0.65 &  0.35 \\
&$0^{\prime++}(2S)$  & 89.7\% & 10.3\% &  7.007 & 0.50 & 0.124 & 
0.49 &  0.47 \\
& $1^{+-}(1S)$ &  100\% &   0\% & 6.481 &  1.17 & 0.580 & 0.48 & 0.37  \\
& $1^{+-}(2S)$ &  100\% &   0\% & 6.954 & 0.17 & 0.045 & 0.61 & 0.44  \\
& $1^{+-}(3S)$ &  100\% &   0\% & 7.024 & 0.12 & 0.029 & 0.66 &0.42   \\
& $2^{++}(1S)$ & 100\% & 0\% & 6.502 &  0.88 & 0.243 & 0.49 & 0.39  \\
& $2^{++}(2S)$ & 100\% & 0\% & 6.917  &   0.12 &  0.033 & 0.55 & 0.60  \\
& $2^{++}(3S)$ & 100\% & 0\% &7.030 &   0.11 & 0.026 & 0.64 & 0.46  \\
 \hline
 \multirow{10}{*}{Ref.~\cite{Lu:2020cns}~} & $0^{\,++}(1S)$  & 38.1\% & 61.9\% & 6.435  & 5.82 & 3.076 & 0.43 & 0.27  \\
 &$0^{\prime++}(1S)$  & 61.9\% & 38.1\% &  6.542 & 6.01  & 2.738  & 0.42 & 0.28  \\
  &$0^{\,++}(2S)$  & 25.0\% & 75.0\% &   6.849  &  0.27 & 0.081 &  0.63 & 0.35  \\
   &$0^{\prime++}(2S)$  & 75.0 \% & 25.0\% & 6.940  & 0.36 & 0.096 & 0.53 &  0.45 \\
    &$0^{\,++}(3S)$  & 44.0\% & 56.0\% &  7.025 & 0.32 & 0.077 & 0.64 & 0.32  \\
     &$0^{\prime++}(3S)$  &56.0 \% & 44.0\% & 7.063  & 0.32 & 0.073  & 0.63 & 0.33  \\
& $1^{+-}(1S)$ &  100\% &   0\% & 6.515 &  6.91 & 3.268 & 0.39 & 0.31  \\
& $1^{+-}(2S)$ &  100\% &   0\% & 6.928 &0.54 & 0.147  & 0.47 & 0.50  \\
& $1^{+-}(3S)$ &  100\% &   0\% & 7.052 & 0.37 & 0.086 &  0.59 &  0.36 \\
& $2^{++}(1S)$ & 100\% & 0\% & 6.543 &  6.29 & 2.862 & 0.39 & 0.32  \\
& $2^{++}(2S)$ & 100\% & 0\% & 6.948 & 0.48  & 0.127 & 0.47 & 0.52  \\
& $2^{++}(3S)$ & 100\% & 0\% &  7.064 & 0.31  &  0.071 & 0.60 & 0.37   \\
 \hline
\end{tabular}
\end{ruledtabular}
\end{table}

\bibliography{tetra_prod}

\end{document}